\documentclass[dvips,12pt]{article}
\usepackage{a4wide, amsmath, latexsym, epsfig,
  psfrag,graphicx, rotating, fancyhdr, tabularx, afterpage,booktabs}
\usepackage{subfigure}
\usepackage{longtable}
\usepackage{pstricks,epsfig,amssymb}
\usepackage[affil-it]{authblk}

\newcommand\ba{\begin{eqnarray}}  
\newcommand\ea{\end{eqnarray}}

\definecolor{Black}{named}{black}
\definecolor{Red}{named}{red}
\definecolor{Blue}{named}{blue}
\definecolor{Green}{named}{green}
\definecolor{Maroon}{named}{brown}
\definecolor{Brown}{named}{brown}
\definecolor{Orange}{named}{orange}

\begin{document}
\begin{titlepage}
\hskip 12 cm {IFJPAN-IV-2015-12}

\begin{center}
{\bf Study of variants for Monte Carlo generators of $\tau\to 3\pi\nu$ decays.}
\end{center} 

\vskip 1cm

\centerline{\bf \Large Z. Was and J. Zaremba }
\vskip 3mm
\centerline{\it Institute of Nuclear Physics, PAN, Krak\'ow, ul. Radzikowskiego 152, Poland}



\abstract{

Low energy QCD (below 2 GeV) is a region of resonance dynamics, 
sometimes lacking satisfactory description 
as compared to precision of available experimental data. 
Hadronic $\tau$ decays offer a probe for such energy regime. 
In general, predictions for decays are model dependent, 
with parameters fitted to experimental results.
Parameterizations differ by amount of 
assumptions and theoretical requirements taken into account. 
Both model distributions and acquired data samples used for fits are results of complex effort.

In this paper, we investigate main parameterizations of $\tau$ decay matrix elements 
for the one- and three-prong channels of three-pion $\tau$ decays. 
Differences in analytical forms of the currents and resulting 
distributions used for comparison with the experimental data are studied.
We use invariant mass spectra of all possible 
pion pairs and the whole three-pion system. 
Also three-dimensional histograms 
spanned over all distinct squared invariant masses are used 
to represent results of models and experimental data.

We present distributions 
from {\tt TAUOLA} Monte Carlo generation and semi-analytical calculation.
These are necessary steps in development for fitting in as model-independent way as possible, 
and to explore multi-million event experimental data samples. 
This includes response of distributions to model variants, 
and/or numerical values of parameters.
Interference effects of currents parts are also studied.

For technical purposes, weighted events are introduced. 
Even though we focus on $3\pi\nu_\tau$ modes, technical aspects of
our study are relevant for all $\tau$ decay modes into three hadrons. 

}

\vfill
{IFJPAN-IV-2015-12}

August 2015

\vspace*{1mm}
\footnoterule
\noindent
{\footnotesize \noindent $^{\dag}$
This project is financed in part from funds of Polish National Science
Centre under decision DEC-2011/03/B/ST2/00107.
}

\end{titlepage}

\section{Introduction}\label{sec:Intro}

One aspect of phenomenological work is to represent theory 
and experimental efforts in form of Monte Carlo generators. 
Then, user should be able to obtain any distribution of interest. 
All details of the detector response can be taken into account if needed.
Introduction to the tasks of modeling with the help of 
Monte Carlo of low energy hadronic interactions can be found, e.g., in \cite{Actis:2010gg},
and will not be repeated here. 
We will concentrate on $\tau$ lepton decays.

Let us point out that
there are various aspects of the work for construction of Monte Carlo, which need to be combined.
On the theoretical side, assumptions have to be translated into
distributions, which can be confronted with the data. 
This task is easy, if underlying theory is well-established and its 
calculation methods based, e.g., on perturbation expansions, are converging sufficiently fast.
For theorists, it would be ideal if experimental results were 
represented in a form of background-subtracted and detector corrected distributions, 
sufficient to constrain all constants (masses, widths, couplings) 
introduced by the theory (or model) used as a basis for the calculation. 

In practice, such a task is often far from being straightforward. 
Assumptions behind theoretical calculations are often not well-established and 
the number of phenomenological constants introduced 
may be too large to be constrained by experimental data.

From the point of view of Monte Carlo techniques, generating processes such as 
$\tau$ decays into 3$\pi \nu$ is rather simple. 
However, necessary for that purpose, hadronic 
currents span distributions over eight-dimensional space 
and are the result of a massive effort. 
Taking into account Lorentz invariance and properties 
of weak couplings of $\tau$-lepton to intermediate virtual $W$ boson,
predictions still require {\it four} complex scalar functions $F_i$ of {\it three} variables each 
(see the definition later in the text).
Such parameterization of matrix elements is commonly used
by $\tau$ decays Monte Carlo generators, e.g., by {\tt TAUOLA};
and since its beginning \cite{Jadach:1990mz, Jadach:1993hs}.

In principle, all properties of the matrix element can be constrained 
from experimental data. All $F_i$ can be fitted in a model independent way as proposed in \cite{Mirkes}. 
In practice, this is highly non-trivial.
Even if necessary for that purpose, direction of $\tau$ neutrino can be 
reconstructed, one has to measure from the data at least {\bf 7} distributions over 
three dimensions with sufficient detail. 
So far, this was never achieved without partial integration 
over some of the phase-space dimensions. 
This practical aspect of the phenomenology work 
should be separated from the previous two 
(theoretical and experimental ones), and 
at the same time offer comfortable and flexible usage.

In this paper, we  investigate differences and physics motivations for 
the presently used in {\tt TAUOLA} hadronic currents of the $\tau \to 3\pi \nu_\tau$ decay channels. 
Let us recall from Ref.~\cite{Shekhovtsova:2015wsa} comparisons of different parameterizations 
for case of the decay mode $\tau \to 2\pi \nu_\tau$. 
In that case, all results are encapsulated in a single one-dimensional distribution.
Our analysis, as we will see, must rely at least on three-dimensional distribution. 
We compare its efficiency to constrain model parameters 
first with the method when one-dimensional histograms are used only. 
To evaluate full size of the model differences, 
integral of $|wt-1|= \big|\frac{|M_{model}|^2}{|M_{model'}|^2}-1\big|$ 
over event samples will be used.
$M_{model}$, $M_{model'}$
denote matrix elements for the two compared model variants.
For completeness, we study the case when 
only total width of the channel is used to constrain the model.  



Theoretical principles embedded in the models and remaining freedom for introducing 
new couplings and/or intermediate resonances has to be studied at the same time. 
We have to keep in mind that additional contributions to the currents can 
modify shapes of the distributions in an nonphysical/uncontrolled 
way and in unexpected regions of the phase-space.
This is especially important when dimensionality 
of experimental distributions is smaller than that of the decay phase-space. 

The paper is organized as follows. In Section~\ref{sec:list}, we present the 
list of currents used for evaluation of distributions. 
Detailed description is delegated to 
Sections \ref{sec:Math}, \ref{sec:CLEO}, \ref{sec:BaBar}, and \ref{sec:rchl}; 
analytic forms of the currents are given. 
In Section \ref{sec:modeling}, we present basic features and motivations behind models used in data fitting.
Section \ref{sec:datarep} consists of introduction to numerical comparisons 
and commonly used representations of experimental data employed in fits.
Section \ref{sec:vsPaper} is devoted to models used by the CLEO collaboration.
In the subsections, numerical results and comparisons are presented.
Section \ref{sec:variants} describes further variants of currents 
that were never used outside the experimental collaboration 
but their existence is mentioned in Ref.~\cite{Asner:1999kj}.
Two main CLEO models are then compared with the BaBar model in Section \ref{sec:vsBaBar}. 
Section \ref{sec:vsRChL} provides comparisons between default 
{\tt TAUOLA} model and two models based on the Resonance Chiral Lagrangian 
(RChL) \cite{Ecker:1989yg}.
For this task, we use mainly one-dimensional histograms; 
results of measurements in such a form are available not only from the CLEO but from the BaBar as well. 
Such distributions were used to constrain model parameters in many of our studies.
Only CLEO used the three-dimensional distributions as an input for fits. 

In Section~\ref{sec:interf} contributions from different parts of currents 
and their interferences are investigated.
We discuss possible consequences of using model assumptions 
and limited data samples in Section~\ref{sec:Properties}.
Summary, Section~\ref{sec:summary}, closes the paper.
In Appendix~\ref{sec:analytical} we present arrangements for 
semi-analytical calculations which can be used for fits into three-dimensional histograms.
In Appendix~\ref{sec:vsPythia} numerical comparison of results from different models 
which are of smaller importance are given. 
Lengthy histograms of three-dimensional form, are given in Appendix~\ref{sec:3dplots}.

\section{List of currents}\label{sec:list}

In {\tt TAUOLA} Monte Carlo generator \cite{Golonka:2000iu, Golonka:2003xt} multiple 
versions of hadronic currents for the two channels of $\tau \to 3\pi \nu_\tau$ decay are available. 
They are the result of theorists effort but also of extensive work of the experimental collaborations. 
Over time, currents became available for use outside of the experiments. 
Unfortunately, sometimes these currents are poorly documented.
We plan to explicitly present differences in formulae and values 
of numerical constants for the main currents available for {\tt TAUOLA}. 
For others, we will skip some details as they are 
of lesser importance or are documented in quoted publications.

The following variants of the currents will be discussed:

\begin{enumerate}
\item {\bf CLEO publ.} -
current described in ref. \cite{Asner:1999kj}. 
This parameterization was developed for data of the $\tau \to \pi^0 \pi^0 \pi^- \nu_\tau$ decays.
Details will be discussed in Sections~\ref{sec:CLEO} and~\ref{sec:vsPaper}.

 
\item {\bf TAUOLA CLEO} - current implemented in {\tt TAUOLA}. 
Code of this current was obtained from the CLEO collaboration and 
distributed with {\tt TAUOLA} (all its versions since Ref.~\cite{Golonka:2000iu}).
See Sections~\ref{sec:CLEO} and~\ref{sec:vsPaper} for details.

\item {\bf TAUOLA CLEO isospin intricate} 
- this is a variant where isospin rotation from the
$\pi^0 \pi^0 \pi^-$ to the $\pi^- \pi^- \pi^+ $ channel was performed for intermediate resonances.
Numerical constants remained unmodified. 
Code was obtained from the CLEO collaboration and 
distributed with {\tt TAUOLA} (all versions since \cite{Golonka:2000iu}) but it 
was never active\footnote{
This parameterization is missing validation with experimental publication; 
only conference contributions \cite{Shibata:2002uv, Schmidtler:1999ur} announce its existence,
while PhD thesis \cite{Hinson:2001cc} fully describes even more elaborated variant.}. 
See Sections~\ref{sec:CLEO} and~\ref{sec:vsPaper} for details.

\item {\bf TAUOLA BaBar} - parameterization numerically equivalent 
to the one used in the BaBar collaboration for basic simulations. 
Details are given in Sections~\ref{sec:BaBar} and~\ref{sec:vsBaBar}.

\item{\bf TAUOLA RChL 2012} - version of RChL model introduced in {\tt TAUOLA} in 2012. 
This model was based on theoretical consideration of that time and fits to invariant mass 
distributions of $\pi^-\pi^-\pi^+$ and $\pi^-\pi^+$ systems \cite{Dumm:2009va}.
For details see Sections~\ref{sec:rchl} and~\ref{sec:vsRChL}.

\item {\bf TAUOLA RChL} - this is the model motivated by 
further theoretical consideration, and comparison with 
experimental data also for $\pi^-\pi^-$ invariant mass, Ref.~\cite{Nugent:2013hxa}.

\item {\bf TAUOLA CPC} - outdated parameterization of Monte Carlo generator. 
Foundation of many technical benchmarks for phase-space 
and other verifications which are documented in \cite{Jadach:1993hs}. 
Also, test distributions documented in Ref.~\cite{Mirkes} are available.

\item {\bf Pythia CLEO} - the parameterization of Ref.~\cite{Asner:1999kj} as 
implemented in {\tt Pythia} \cite{Sjostrand:2014zea}. 
Current nearly identical to the {\bf TAUOLA CLEO isospin intricate} is used. 
Currents of {\tt Pythia} differ only by numerical constants. 
In the present paper, 
we use numerical results of {\tt Pythia} version 8.201. 
This current will be discussed in Appendix~\ref{sec:vsPythia}

\end{enumerate}

At present, {\bf TAUOLA CLEO} is the default choice which should give the same result 
regardless of which version of {\tt TAUOLA} is used. 
It remained unmodified for at least 14 years. 
It is identical for {\tt FORTRAN} and {\tt C++} implementations.
Differences in distributions between {\bf CLEO publ.}, {\bf TAUOLA CLEO}, 
{\bf TAUOLA CLEO isospin intricate} and 
{\bf Pythia CLEO} are presented with all technical and numerical aspects. 
For completeness, we present numerical results from {\bf TAUOLA RChL}.
This last current is aimed to represent data with good precision, 
and at the same time improve relation with present day theoretical calculations. 
Numerical results for other variants of the currents are given in less detail or 
are not presented at all.

\section{Common mathematical functions}\label{sec:Math}

All currents parameterizations use the same basic formulae.
Equation \eqref{eq:Hadcur}
defines\footnote{We use five form factors instead of four imposed by 
Lorenz invariance for practical purpose.
In principle $F_3$ can be represented as linear combination of
contributions to $F_1$ and $F_2$.} 
hadronic current. It is common for all three-scalars decay channels of {\tt TAUOLA} 
and  with its help matrix element of $\tau$ decays can be calculated in a straightforward manner  
\cite{Jadach:1990mz, Jadach:1993hs},

{\footnotesize
\begin{eqnarray}
J^\mu &=N &\bigl\{T^\mu_\nu \bigl[ c_1 (p_2-p_3)^\nu F_1 + c_2 (p_3-p_1)^\nu
 F_2 + c_3 (p_1-p_2)^\nu F_3 \bigr]\nonumber\\
& & + c_4 
q^\mu F_4  -{ i \over 4 \pi^2 F^2}      c_5 \epsilon^\mu_{.\ \nu\rho\sigma} p_1^\nu p_2^\rho p_3^\sigma
 F_5      \bigr\},
\label{eq:Hadcur}
\end{eqnarray}
}

where $T_{\mu\nu} = g_{\mu\nu} - Q_\mu Q_\nu/Q^2$ denotes the transverse
projector, and $Q^\mu=(p_1+p_2+p_3)^\mu$ 
is the momentum of the hadronic system\footnote{In {\tt TAUOLA} code defined as {\tt PAA(4)}.}.  
The decay products are ordered $\pi^- \pi^- \pi^+ $ for the three-prong channel and $\pi^0 \pi^0 \pi^- $ 
for the one-prong channel and the pions four-momenta\footnote{
In {\tt TAUOLA} code represented as {\tt PIM1(4), PIM2(4), PIM3(4)}.}  
are denoted as $p_1$, $p_2$ and $p_3$, respectively.
The same ordering is used in further equations also for masses ($m_i$).
The $\epsilon^\mu_{.\ \nu\rho\sigma}$ is the Levi-Civita symbol\footnote{
The $\epsilon^\mu_{.\ \nu\rho\sigma} p_1^\nu p_2^\rho p_3^\sigma$ 
is coded in subroutine PROD5.}.
In equations of this and the following sections we use notation: 
$s_i=(p_j+p_k)^2$ where i $\neq$ j $\neq$ k $\neq$ i. 
Constants: $c_1$, $c_2$, etc. are Clebsch-Gordan coefficients, 
defined specifically for particular hadronic current used.

Eqs. \eqref{eq:momentum}, \eqref{eq:gamma} and \eqref{eq:BW} describe Breit-Wigner 
functions that are later used in definition of form factors:

{\footnotesize
\begin{equation}
 P(S, m_1, m_2) \ = \ 
 \frac{\sqrt{(S-(m_1 + m_2)^2)(S-(m_1 - m_2)^2)}}{\sqrt{S}},
\label{eq:momentum}
\end{equation}

\begin{equation}
 \Gamma_{L-wave} (S, M, \Gamma, m_1, m_2, L) \ = \ 
 \Gamma\frac{M}{\sqrt{S}}\bigg( \frac{P(S, m_1, m_2)}{P(M^2, m_1, m_2)} \bigg)^{2L+1},
\label{eq:gamma}
\end{equation}

\begin{equation}
BW(S, M,\Gamma, m_1, m_2, L) \ = \
\frac{M^2}{S - M^2 - i M \Gamma_{L-wave} (S, M, \Gamma, m_1, m_2, L)}.
\label{eq:BW}
\end{equation}
}

These are typical building blocks useful for hadronic currents parameterizations, such as of
Gounaris-Sakurai parametrization \cite{Gounaris:1968mw} for $\rho \to \pi \pi $. 

\section{Hadronic current in models of the CLEO category} \label{sec:CLEO}

Let us describe the default, 
hadronic currents of the CLEO modeling - {\bf TAUOLA CLEO}.
In eq. \eqref{eq:7b}, \eqref{eq:8b} the analytic form of the hadronic 
current for the $\tau \to 2\pi^0\pi^-\nu_\tau$ decay channel is presented\footnote{
Constants: $\beta, \beta_1, \beta_2$ 
etc. are coded in {\tt TAUOLA} as BET, BT1, BT2, etc.}.
Its functional form is extracted from {\tt TAUOLA} 
code\footnote{Such approach checks if over the years any modifications were (intentionally or not) introduced.} 
but has a form exactly as described in appendix A of Ref.~\cite{Asner:1999kj}. 
Notations differ slightly. This form of hadronic current is also used for the 
$\tau \to 2\pi^-\pi^+\nu_\tau$ decay channel in default {\tt TAUOLA} setup. 
It will be discussed later why such a choice is justified. 
In PDG tables Ref.~\cite{Agashe:2014kda}, some mesons have the same name but different mass.
In our paper;
$a_1(1260)$, $\sigma$ (or $f_0$(500)), $f_2(1270)$, $f_0(1370)$, 
$\rho(770)$, $\rho'(1450)$, $K^*(892)$ are used. 
Finally, meson $a_1'(1700)$ as in Ref.~\cite{Asner:1999kj}.
Masses in names will be skipped from this point on.

{\footnotesize
\begin{multline}
F_1 \ = \
\bigg( \frac{M_{a_1}^2}{Q^2-M_{a_1}^2 - \frac{i M_{a_1} \Gamma_{a_1}}{1.3281\cdot0.806} \cdot WGA(Q^2)} 
+\beta \frac{M_{a_1'}^2}{Q^2-M_{a_1'}^2 - \frac{i M_{a_1'} \Gamma_{a_1'}}{1.3281\cdot0.806} \cdot WGA(Q^2)}
\bigg)\cdot \\
\bigg(
    \beta_1 \cdot BW(s_1, M_\rho,\Gamma_\rho, m_2, m_3, 1)+ \beta_2 \cdot BW(s_1, M_{\rho'},\Gamma_{\rho'}, m_2, m_3, 1)\\
    -\beta_3 \cdot \frac{(s_3-m_3^2)-(s_1-m_1^2)}{3} \cdot BW(s_2, M_\rho,\Gamma_\rho, m_3, m_1, 1)\\
    -\beta_4 \cdot \frac{(s_3-m_3^2)-(s_1-m_1^2)}{3} \cdot BW(s_2, M_{\rho'},\Gamma_{\rho'}, m_3, m_1, 1)\\
    +\beta_5 \cdot \frac{(Q^2+s_3-m_2^2)(2m_3^2+2m_1^2-s_3)}{18s_3} \cdot BW(s_3, M_{f_2},\Gamma_{f_2}, m_1, m_2, 2)\\
    +\beta_6 \cdot \frac{2}{3} \cdot BW(s_3, M_\sigma,\Gamma_\sigma, m_1, m_2, 0)\\
    +\beta_7 \cdot \frac{2}{3} \cdot BW(s_3, M_{f_0},\Gamma_{f_0}, m_1, m_2, 0)
\bigg),   \\
\label{eq:7b}
\end{multline}
}

$F_2$ has the same functional form as $F_1$. The only difference is interchange 
for its arguments indices 1 and 2 in eq. \eqref{eq:Hadcur}, and that constant $c_2$ has 
opposite sign to $c_1$. Note, that $\beta$ is always set to 0. 
We suspect that it was introduced by the CLEO collaboration for studies of $a_1'$ influence. 

{\footnotesize
\begin{multline}
F_3 \ = \
\bigg( \frac{M_{a_1}^2}{Q^2-M_{a_1}^2 - \frac{i M_{a_1} \Gamma_{a_1}}{1.3281\cdot0.806} \cdot WGA(Q^2)} 
+\beta \frac{M_{a_1'}^2}{Q^2-M_{a_1'}^2 - \frac{i M_{a_1'} \Gamma_{a_1'}}{1.3281\cdot0.806} \cdot WGA(Q^2)}
\bigg)\cdot \\
\bigg(
    \beta_3 \cdot \frac{(s_2-m_2^2)-(s_3-m_3^2)}{3} \cdot BW(s_1, M_\rho,\Gamma_\rho, m_2, m_3, 1)\\
    +\beta_3 \cdot \frac{(s_3-m_3^2)-(s_1-m_1^2)}{3} \cdot BW(s_2, M_\rho,\Gamma_\rho, m_3, m_1, 1)\\
    +\beta_4 \cdot \frac{(s_2-m_2^2)-(s_3-m_3^2)}{3} \cdot BW(s_1, M_{\rho'},\Gamma_{\rho'}, m_2, m_3, 1)\\
    +\beta_4 \cdot \frac{(s_3-m_3^2)-(s_1-m_1^2)}{3} \cdot BW(s_2, M_{\rho'},\Gamma_{\rho'}, m_3, m_1, 1)\\
    -\beta_5 \cdot \frac{(s_1-m_1^2)-(s_2-m_2^2)}{2} \cdot BW(s_3, M_{f_2},\Gamma_{f_2}, m_1, m_2, 2)   
\bigg).   \\
\label{eq:8b}
\end{multline}
}

Form factors $F_4$ and $F_5$ are set to zero\footnote{
Such choice is dictated by absence of scalar (like $\pi$(1300)) 
and vector (like $\omega$) intermediate states in this model.}.
Note, that normalization factor used in $a_1$ and $a_1'$ width is the same.
The $Q^2$ dependence of the $a_1$ width in its propagator is given by the 
formula\footnote{
Not all numerical values in the formula below are coming from fits. 
Explicitly coded numerical constants in {\tt TAUOLA} have clear physical meaning: 
0.1753=$(3m_\pi)^2$, 0.823=$(M_\rho+m_\pi)^2$,
0.1676=$(2m_{\pi^0}+m_\pi)^2$.} \eqref{eq:a1cleo}. 
It was obtained from analysis performed by the CLEO collaboration \cite{Asner:1999kj}. 
Contributions from the three main $a_1$ decay channels are taken into account. 
Complicated form is determined by the decay channels: 
as $a_1$ virtuality gets larger thresholds are crossed, allowing more 
decay channels to open, therefore changing the $Q^2$ dependence of the effective width.  
$Q^2$ is given in $\mathrm{GeV^2}$ units.
In the equation below: $C_{3\pi}=0.2384^2$ and $C_{K^*}=4.7621^2 C_{3\pi}$.

{\footnotesize
\begin{multline}
WGA(Q^2) \ = \ \\ C_{3\pi} \cdot
   \begin{cases}
    0 & \text{if } Q^2 < 0.1753,\\
    5.809(Q^2-0.1753)^3[1-3.0098(Q^2-0.1753)+4.5792(Q^2-0.1753)^3)] & \text{if } 0.1753<Q^2<0.823,\\
    -13.914+27.679Q^2-13.393Q^4+3.1924Q^6-0.10487Q^8 & \text{if } Q^2>0.823,\\
   \end{cases} \\
  + C_{3\pi} \cdot \begin{cases}
    0 & \text{if } Q^2 < 0.1676,\\
    6.2845(Q^2-0.1676)^3[1-2.9595(Q^2-0.1676)+4.3355(Q^2-0.1676)^3] & \text{if } 0.1676<Q^2<0.823,\\
    -15.411+32.088Q^2-17.666Q^4+4.9355Q^6-0.37498Q^8 & \text{if } Q^2>0.823,\\
   \end{cases} \\
 + C_{K^*}\cdot
   \begin{cases}
    \frac{\sqrt{(Q^2-(M_{K^*}+m_K)^2)(Q^-(M_{K^*}-m_K)^2)}}{2Q^2} & \text{if } Q^2>(M_{K^*}+m_K)^2, \\
    0 & \text{if } Q^2\leq(M_{K^*}+m_K)^2. \\
   \end{cases} \\
\label{eq:a1cleo}
\end{multline}
}

Let us now turn to the case of {\bf TAUOLA CLEO isospin intricate}. 
It differs from the previous for the $\pi^- \pi^- \pi^+ $ only. 
Formulae \eqref{eq:7b} and \eqref{eq:8b} are still used for  $\pi^0 \pi^0 \pi^- $,
but they are replaced by \eqref{eq:7} and \eqref{eq:8} in the three-prong channel. 
Those are also extracted from the {\tt TAUOLA} code, 
but from part which was not available for the general use. 

{\footnotesize 
\begin{multline}
F_1 \ = \
\bigg( \frac{M_{a_1}^2}{Q^2-M_{a_1}^2 - \frac{i M_{a_1} \Gamma_{a_1}}{1.3281\cdot0.806} \cdot WGA(Q^2)} 
+\beta \frac{M_{a_1'}^2}{Q^2-M_{a_1'}^2 - \frac{i M_{a_1'} \Gamma_{a_1'}}{1.3281\cdot0.806} \cdot WGA(Q^2)}
\bigg)\cdot \\
\bigg(
    \beta_1 \cdot BW(s_1, M_\rho,\Gamma_\rho, m_2, m_3, 1)+ \beta_2 \cdot BW(s_1, M_{\rho'},\Gamma_{\rho'}, m_2, m_3, 1)\\
    -\beta_3 \cdot \frac{(s_3-m_3^2)-(s_1-m_1^2)}{3} \cdot BW(s_2, M_\rho,\Gamma_\rho, m_3, m_1, 1)\\
    -\beta_4 \cdot \frac{(s_3-m_3^2)-(s_1-m_1^2)}{3} \cdot BW(s_2, M_{\rho'},\Gamma_{\rho'}, m_3, m_1, 1)\\
    +\beta_5 \cdot \frac{(s_2-m_2^2)-(s_3-m_3^2)}{2} \cdot BW(s_1, M_{f_2},\Gamma_{f_2}, m_2, m_3, 2)\\
    +\beta_5 \cdot \frac{(Q^2+s_2-m_2^2)(2m_3^2+2m_1^2-s_2)}{18s_2} \cdot BW(s_2, M_{f_2},\Gamma_{f_2}, m_3, m_1, 2)\\
    +\beta_6 \cdot \frac{2}{3} \cdot BW(s_2, M_\sigma,\Gamma_\sigma, m_3, m_1, 0)\\
    +\beta_7 \cdot \frac{2}{3} \cdot BW(s_2, M_{f_0},\Gamma_{f_0}, m_3, m_1, 0)
\bigg). \\   
\label{eq:7}
\end{multline}
}
Again, $F_2$ has the same functional form as $F_1$. The only difference is interchange 
of indices 1 and 2 for its arguments in eq. \eqref{eq:Hadcur}, and that constant $c_2$ has 
opposite sign to  $c_1$. 

{\footnotesize
\begin{multline}
F_3 \ = \
\bigg( \frac{M_{a_1}^2}{Q^2-M_{a_1}^2 - \frac{i M_{a_1} \Gamma_{a_1}}{1.3281\cdot0.806} \cdot WGA(Q^2)} 
+\beta \frac{M_{a_1'}^2}{Q^2-M_{a_1'}^2 - \frac{i M_{a_1'} \Gamma_{a_1'}}{1.3281\cdot0.806} \cdot WGA(Q^2)}
\bigg)\cdot \\
\bigg(
    \beta_3 \cdot \frac{(s_2-m_2^2)-(s_3-m_3^2)}{3} \cdot BW(s_1, M_\rho,\Gamma_\rho, m_2, m_3, 1)\\
    +\beta_3 \cdot \frac{(s_3-m_3^2)-(s_1-m_1^2)}{3} \cdot BW(s_2, M_\rho,\Gamma_\rho, m_3, m_1, 1)\\
    +\beta_4 \cdot \frac{(s_2-m_2^2)-(s_3-m_3^2)}{3} \cdot BW(s_1, M_{\rho'},\Gamma_{\rho'}, m_2, m_3, 1)\\
    +\beta_4 \cdot \frac{(s_3-m_3^2)-(s_1-m_1^2)}{3} \cdot BW(s_2, M_{\rho'},\Gamma_{\rho'}, m_3, m_1, 1)\\
    -\beta_5 \cdot \frac{(Q^2+s_1-m_1^2)(2m_2^2+2m_3^2-s_1)}{18s_2} \cdot BW(s_1, M_{f_2},\Gamma_{f_2}, m_2, m_3, 2)\\
    -\beta_5 \cdot \frac{(Q^2+s_2-m_2^2)(2m_3^2+2m_1^2-s_2)}{18s_2} \cdot BW(s_2, M_{f_2},\Gamma_{f_2}, m_3, m_1, 2)\\
    +\beta_6 \cdot \frac{-2}{3} \cdot BW(s_1, M_\sigma,\Gamma_\sigma, m_2, m_3, 0)\\
    +\beta_6 \cdot \frac{2}{3} \cdot BW(s_2, M_\sigma,\Gamma_\sigma, m_3, m_1, 0)\\
    -\beta_7 \cdot \frac{-2}{3} \cdot BW(s_1, M_{f_0},\Gamma_{f_0}, m_2, m_3, 0)\\
    -\beta_7 \cdot \frac{2}{3} \cdot BW(s_2, M_{f_0},\Gamma_{f_0}, m_3, m_1, 0)
\bigg). \\   
\label{eq:8}
\end{multline}
}

One can ask, why the isospin symmetry was imposed 
taking into account final states (that means distributions only) in the version of the code 
used as default in {\tt TAUOLA} for so many years, even though it is 
not the property of the model used, see eg. Ref.~\cite{Rouge:2003bg}.
We will return to this point later at a time of discussing numerical results.
Current of eq. \eqref{eq:7b} and \eqref{eq:8b} is used in {\tt TAUOLA} 
for one-prong and three-prong of the three-pion channels in most of applications\footnote{
Note, that later the BaBar collaboration was using both 
of these options as alternative to their default parameterization.}. 
Note, that neither Ref.~\cite{Schmidtler:1999ur} nor \cite{Shibata:2002uv} 
give functional form of $\tau \to 2\pi^-\pi^+\nu_\tau$ current, 
also Ref.~\cite{Asner:1999kj} presents functional form of hadronic 
current for $\tau \to 2\pi^0\pi^-\nu_\tau$ only. 
Parameters used in currents, will be presented later.

\section{Analytic form of hadronic current of {\bf TAUOLA BaBar}}\label{sec:BaBar}

This model became public recently, it was developed by the experimental collaboration. 
Mentioned in Section~\ref{sec:Intro}, isospin symmetry and limitation of 
quality of the data was probably behind the choice of form factors used by BaBar as well. 
They are the same for $\tau \to 2\pi^0\pi^-\nu_\tau$ and $\tau \to 2\pi^-\pi^+\nu_\tau$. 
BaBar model does not include resonances $f_0$, $f_2$ and $\sigma$. 
Note, that BaBar choice was introduced after the CLEO currents 
(both variants) were known and available for them.
Equations in this section are extracted from {\tt TAUOLA} 
code numerically equivalent to the one used by BaBar\footnote{
The authors would like to acknowledge Swagato Banerjee and Tomasz 
Przedzinski for cross-validation of our
modifications of the $\tau \to 3\pi \nu$ hadronic current called {\tt 
TAUOLA BaBar}, with the {\tt TAUOLA} installation in BaBar framework
  at a statistical level of 100M events. 
For comparisons standard distributions prepared with 
{\tt MC-TESTER} \cite{Davidson:2008ma} were used.
}. 

{\footnotesize
\begin{multline}
F_1 \ = \
 \frac{M_{a_1}^2}{Q^2-M_{a_1}^2 - i M_{a_1} \frac{GFUN(Q^2)}{GFUN(M_{a_1}^2)}} \\
 \cdot
   \begin{cases}
    BW(s_1, M_\rho,\Gamma_\rho, m_2, m_3, 1)+\frac{\beta_1}{1+\beta_1}BW(s_1, M_{\rho'},\Gamma_{\rho'}, m_2, m_3, 1) & \text{if } s_1>(m_\pi+m_\pi)^2, \\
    \frac{M_\rho^2}{M_\rho^2-s_1}+\frac{\beta_1}{1+\beta_1}\frac{M_{\rho'}^2}{M_{\rho'}^2-s_1} & \text{if } s_1\leq(m_\pi+m_\pi)^2, \\
   \end{cases}
\end{multline}}
\\
$F_2(s_2)$ coincide with $F_1(s_1)$. 
It has an opposite sign as well;
$F_3$ = 0; $F_4$ = 0; $F_5$ = 0.

{\footnotesize
\begin{equation}
 GFUN(x) =
  \begin{cases}
   4.1(x-9m_{\pi_0}^2)^3[1-3.3(x-9m_{\pi_0}^2)]+5.8(x-9m_{\pi_0}^2)^2 & \text{if } x < (M_\rho + m_\pi)^2, \\
   x(1.623+\frac{10.38}{x}-\frac{9.32}{x^2}+\frac{0.65}{x^3})       & \text{if } x \geq (M_\rho + m_\pi)^2. \\
  \end{cases}
\label{eq:a1babar}
\end{equation}
}

Note, that eqs. \eqref{eq:a1babar} and \eqref{eq:a1cleo} aim at $a_1$ phase-space dependence, 
but eq. \eqref{eq:a1babar} does so in simpler manner. 
CLEO model uses more sophisticated polynomial interpolation with inclusion of 
$K^*K$ production threshold; above $a_1$ can decay relatively abundantly into pair of kaons and a pion.

\section{Resonance Chiral Lagrangian currents} \label{sec:rchl}

Our first hadronic current based on the Resonance Chiral Lagrangian scheme - {\bf TAUOLA RChL 2012} 
was prepared for Ref.~\cite{Shekhovtsova:2012ra}. 
Similarly to parameterization of BaBar, the same form of current was used 
for one- and three-prong $\tau$ decay modes into three pions. 
As an input for fits, only two, one-dimensional histograms of BaBar data were used. 
Later on, the invariant mass of $\pi^-\pi^-$ pair became available. 
It turned out that contributions which, 
under isospin symmetry, 
provide different form of the current for one- and three-prong decay modes was useful. 
We refer to this later parameterization as {\bf TAUOLA RChL}; 
it is documented in Ref.~\cite{Nugent:2013hxa}. 

We will return to isospin symmetry aspect, 
when we will discuss sensitivity of different distributions to specific parts of current contributions. 
This may hint, why isoscalar terms introduced by CLEO many years ago, were actually dropped out
in currents, such as {\bf TAUOLA BaBar} and {\bf TAUOLA RChL 2012}. 

\section{Practical aspects of model construction} \label{sec:modeling}

There are two main paths to be taken in model building. 
Empirical approach is to prepare something describing the data in the best possible way, 
without considering how physical process works. 
Other choice is to start from theoretical principles, obtain distributions
and introduce only necessary adaptations at a time of comparison with data. 
Of course, merging those two approaches is desirable and 
even essential. There are unquestioned rules like Lorentz invariance which have to be obeyed.
All models used in {\tt TAUOLA} and in our paper, respect such rules. 

In particle physics we expect Breit-Wigner 
distributions to describe shapes of experimental distributions. 
One should bear in mind that precision of experimental data of today can be very high. 
Multiple millions of events were collected for individual $\tau$ decay channels.  
On the other hand, once inspection of the assumptions behind theoretical
models is performed, rather modest estimations for systematic
uncertainties of the predictions are obtained \cite{Roig:2012zj}. 
For example, for RChL models, in Ref.~\cite{Shekhovtsova:2012ra} Section 7.4
it was argued that on the basis of theoretical arguments alone, 
the predicted uncertainty can in principle be as high as 30 \%.  
Even behind such a basic principle like isospin symmetry uncertainty at 
the level of 5-10 \% can be expected at least in some cases, 
see, e.g., Ref.~\cite{Actis:2010gg} Section 5.11. That is again more than one order 
of magnitude worse than precision of the experimental data. 
Comparing to differences in results of Section~\ref{sec:numCLEO} 
it is about the same order of magnitude.

That is why, comparison of model's results with the data can be an exciting 
source of inspiration for further model developments, even if attained
agreement with the data is not perfect at the beginning. 
Such development can be particularly promising for the models precisely representing 
distributions of several $\tau$ decay channels and/or final states in $e^+e^- \to hadrons$.

In case of {\bf TAUOLA CLEO}, at start, model of Ref.~\cite{Kuhn:1990ad} was used. 
With time, as a consequence of comparisons with the data, it evolved
into form where Breit-Wigner resonances seen in experimental data were added.
Some of the model assumptions, in particular numerical values of the chiral 
couplings in soft limit were compromised. One should keep in mind that
only at very low energies chiral dynamics is expected to determine 
unambiguously the hadron form factors probed by semileptonic $\tau$ decays 
\cite{Weinberg:1978kz, Gasser:1983yg, Colangelo:1996hs}.

At present, more elegant approach is taken; the RChL model. 
It was started from theoretical calculations accounting 
to expansion in number of colors $ N_c \to \infty$ \cite{'tHooft:1973jz, Witten:1979kh}.
Later, this model required empirical correction by the 
$\sigma$ meson introduced in Ref.~\cite{Nugent:2013hxa} as an additional Breit-Wigner. 
Authors are still working out a way to include $\sigma$ 
and possibly other isoscalars within RChL scheme using 
unitarity and analiticity constraints to include this re-scattering effect.
  
Other interesting approach in modeling was taken, e.g., in Ref.~\cite{Bondar:2002mw}.
It describes hadronic current for $\tau \to 4\pi\nu_\tau$. 
This modeling is in part data driven. 
As measurements of $e^+ e^-$ annihilation into four pions arrived ~\cite{Akhmetshin:1998df}, 
they could be used to model distribution of invariant mass of the whole $4\pi$ system\footnote{
It was done by relating $2\pi^+ 2\pi^-$ and $\pi^+ \pi^- 2\pi^0$ systems 
produced in $e^+ e^-$ annihilation with $\pi^+ 2\pi^- \pi^0$ and $\pi^- 3\pi^0$ produced in $\tau$ decays.
For that, isospin symmetry was applied to distributions measured from $e^+ e^-$ annihilation 
giving prediction on shapes of four-pion system invariant mass spectra in $\tau$ decays.}. 
Authors introduced correction into matrix element to reproduce experimental distribution. 
Then, $a_1 \pi$ dominance as intermediate state was assumed. 
From that point on, any desired modeling of $a_1$ could be used. 
The one of Ref.~\cite{Bondar:1999ac} was chosen as it was applied for $e^+ e^-$ data as well. 

Those are only a few of many possible approaches. 
There is no straightforward way of telling that one is better than other.
Actual choice has to be motivated by quality and precision of experimental data. 
Let us address this point in the next section.

\section{Data representations and numerical methods} \label{sec:datarep}

Experimentalists and theoreticians may prefer distinct forms for 
data representation in model building, comparisons and fits. 
In this section we will concentrate on methods we use and those that were 
used for constraining investigated models in the past.

The most simple way to compare models and also to fit the experimental data 
is to look only at the total width of the channel\footnote{
{\tt Option 1} was used, e.g., in Ref.~\cite{Kuhn:2006nw} for 5$\pi$ currents.} 
(we will call it {\tt option 1}).
Such approach is used when there is no sufficient data available for a more advanced fit. 
Even though it is only one number, it can be used as a first test of particular model.
Total width is usually well documented by PDG tables, see, e.g., Ref.~\cite{Agashe:2014kda} 
including careful analysis of experimental statistical and systematical error.

Next approach is to look at one-dimensional distributions. 
In this case, one should seek the possibility of having experimental data 
for all possible invariant mass distributions. 
This is not always possible. 
Using as an example the $\pi^-\pi^-\pi^+$ channel, one must be aware that 
it is easy to have many models that fit perfectly, let's say the $\pi^-\pi^-\pi^+$ spectra, 
while having distributions of $\pi^-\pi^-$ and $\pi^-\pi^+$ off. 
We can articulate 3 options of one-dimensional distributions 
in use for the case of $\tau \to 3\pi\nu_\tau$ current input:
\begin{itemize}
\item $\pi^-\pi^-\pi^+$ system invariant mass only\footnote{
Similar to {\tt option 2} fit was performed for 4$\pi$ current in Ref.~\cite{Bondar:2002mw} 
with only invariant mass of four pions used.} ({\tt option 2}).
\item $\pi^-\pi^-\pi^+$ and $\pi^-\pi^+$ invariant masses\footnote{
{\tt Option 3} was used in fits of {\bf TAUOLA RChL 2012}. 
Probably also BaBar model was fitted in this way.} ({\tt option 3}).
\item $\pi^-\pi^-\pi^+$, $\pi^-\pi^+$ and $\pi^-\pi^-$ invariant masses\footnote{
{\tt Option 4} was used for fits of {\bf TAUOLA RChL}.} ({\tt option 4}).
\end{itemize}

In more advanced approach of {\bf Dalitz distributions} ({\tt option 5}) 
slices in $Q^2$ were used already in Ref.~\cite{Asner:1999kj}. 
This way of data representation is in nature three-dimensional\footnote{ 
There are 8 plots, each in different range of $Q^2$; on x-axis there is $s_1$ and $s_2$ is on y-axis. 
$s_1$ is always chosen as the bigger of the two possible $\pi^-\pi^+$ pair invariant masses. 
See plots of appendix \ref{sec:3dplots}.}.
Models originating from the CLEO collaboration were fitted in that way.
To explore the whole structure of the decay channel methods of 
Ref.~\cite{Mirkes} have to be introduced ({\tt option 6}). 
Up to now, no complete analysis was performed with their help. 
Only Ref.~\cite{Asner:1999kj} mentions cross-check with such methods.

Let us turn to practical aspects of our paper comparisons 
for distributions with options listed above. 
For that purpose we describe re-weighting method.
We use it often due to its numerical efficiency and simplicity. 
When generating events from Monte Carlo generator for different models, 
statistically independent event samples are produced.
But one may ask what is the probability of getting particular 
event with certain set of momenta depending on the model we use.
Having all momenta of particles in an event, 
one can calculate the amplitude for such specific configuration. 
Therefore, one can calculate amplitudes for the event using each model and compare. 
This leads to the possibility of re-weighting events by 
using one model as reference and instead of generating a new sample with a different model, 
calculating amplitudes from that other model for existing events. 
Ratio of new amplitude squared to old one, can be then attributed to the event as its weight. 
Thanks to weights use, samples are correlated. 
Statistical error affects the difference for the compared models only.
Using this strategy for a different model we may not only get the new distributions with 
different shape but also its integral, that is the total width of the channel. 
Method can be applied to minuscule changes of model parameters,
which are needed, e.g., for calculation of derivatives.

Re-weighting events can be used to estimate the contribution from parts of the model amplitude as well. 
The ratio of investigated part of the amplitude squared 
to a whole amplitude squared provides then the weight. 
Evaluation of interferences between parts of currents is also possible on the event by event basis.
One has to evaluate how far from 1 is the sum of weights coming from all parts.

Let us look into that in more detail.
For any integrated distributions  
bin always represents average weight of events filling it. 
Once histogram is filled, information from separate events is lost. 
Because of that, interference observed in any histogram is reduced with respect to its full size.
Effect in integrated distribution is dependent on selection of bins.
By comparing interference coming from the event by event analysis and integrated distributions 
one may evaluate how viable used distributions are to constrain interference 
and if model can be explored with available data sufficiently well.
Integrated contributions from parts of current are identical for any distribution, 
but interferences may differ vastly. 
Analysis of option 1 is expected to give very poor insight on interferences,
and options 2-4 can give only limited information.
Option 5 is expected to give the best way to explore models in that regard. 
Yet, it still does not explore the whole structure of multidimensional phase-space, 
contrary to methods of Ref.~\cite{Mirkes}.   

\section{{\bf TAUOLA CLEO}, {\bf TAUOLA CLEO isospin intricate},
{\bf CLEO publ.}}\label{sec:vsPaper}

Beside functional form of currents, numerical parameters affect shapes of distributions. 
Among them, mainly masses, widths, and couplings of the intermediate resonances differ. 
That is why, we collect all constants to the last detail,
taking into account variants in the implementations.
The set of parameters\footnote{Note: we present here only those 
parameters that are different as only those can introduce any
discrepancies between {\bf CLEO publ.} and {\bf TAUOLA CLEO}.
The complete list of parameters is given in Section~\ref{sec:vsBaBar}.} 
is extracted from the {\tt TAUOLA} code, 
to make sure that they were used in {\bf TAUOLA CLEO} 
and {\bf TAUOLA CLEO isospin intricate} currents. These currents rely on analysis of data 
using {\tt option 5}, but further cross-checks using {\tt option 6}, 
were also mentioned in CLEO publications. 
Published experimental data 
for $\tau \to \pi^0\pi^0\pi^-\nu_{\tau}$ were used by the collaboration. 
Predictions for $\tau \to \pi^-\pi^-\pi^+\nu_{\tau}$ were documented only in conference contributions 
\cite{Shibata:2002uv, Schmidtler:1999ur}.
It is not the theoreticians or MC authors right to decide whether predictions for the $\pi^-\pi^-\pi^+$
channel should be obtained from the $\pi^0\pi^0\pi^-$ case 
at the level of distribution or taking into account isospin rotation of intermediate states.
Actual choice must rely on analysis of systematic errors.
Our default current {\bf TAUOLA CLEO} implies that one considers 
contributions from intermediate states: $f_0$, $f_2$, $\sigma$ as effective ways to improve distributions. 
We use then distribution level isospin symmetry to get predictions for $\pi^-\pi^-\pi^+$
from $\pi^0\pi^0\pi^-$; we use the same current for the two cases. 
Second variant, {\bf TAUOLA isospin intricate}, 
was also prepared by the collaboration but was not published. 
We have nonetheless distributed it together with all versions of {\tt TAUOLA}, 
but not made available for direct {\tt TAUOLA} use\footnote{
There are other options of hadronic currents distributed, 
but never made available for general use as well. 
The aim was to stimulate future work 
or for archivization. 
That was for example the case of $\pi^-\pi^-\pi^+\pi^0$ current used for background modeling 
in Delphi analysis of Ref.~\cite{Abreu:1998cn} 
and $(K\pi\pi)^-$ currents of OPAL~\cite{Abbiendi:1999cq}.
Alternative ALEPH parameterization of hadronic currents was documented in 
Refs.~\cite{Golonka:2000iu, Golonka:2003xt}.}. 
To activate it, modification of code was always necessary. 
Let us stress that the numerical difference between the two options is comparable/smaller 
than the differences to other parameterizations presented in later sections. 
We believe also that the choice lies within the systematic uncertainties. 
This would of course have changed with CLEO publication 
or with the future publications of other collaborations. 
For further discussion and numerical comparison of different current options 
see the following subsections and Sections~\ref{sec:vsBaBar}, \ref{sec:vsRChL}.

The model described in Ref.~\cite{Asner:1999kj} and the implementation used in {\tt TAUOLA} 
differ only by parameters stored in Table~\ref{tab:vsCLEO}. 
One should stress that Ref.~\cite{Asner:1999kj} is for the $\pi^0\pi^0\pi^-$ mode, 
whereas {\bf TAUOLA CLEO} use the same current for the $\pi^-\pi^-\pi^+$ mode as well.

\begin{table}[h]
\centering
\begin{tabular}{|c|c|}
\hline
TAUOLA CLEO & CLEO of Ref.~\cite{Asner:1999kj} \\ \hline 
$M_{\rho}$ = MRO =0.7743 GeV & $M_{\rho}$ = MRO =0.774 GeV \\
$\Gamma_{\rho}$ = GRO =0.1491 GeV & $\Gamma_{\rho}$ = GRO =0.149 GeV \\
$M_{\tau}$ = 1.777 GeV & not mentioned in the paper \\
$M_{a_1}$ = 1.275 GeV & $M_{a_1}$ = 1.331 GeV \\
$\Gamma_{a_1}$ = 0.700 GeV & $\Gamma_{a_1}$ = 0.814 GeV \\
\hline
\end{tabular}
\caption{Parameters differing in {\tt TAUOLA} and Ref.~\cite{Asner:1999kj}. 
These differences in parameters are the reason of differences visible in Figure \ref{fig:5}. \label{tab:vsCLEO}}
\end{table}

Analytic form  of {\bf TAUOLA CLEO isospin intricate} current for $\tau \to \pi^-\pi^-\pi^+\nu_\tau$ 
can be understood as a simple consequence of isospin symmetry applied to the
$\tau \to \pi^0\pi^0\pi^-\nu_\tau$ case.
But we must make sure if the hadronic current 
to be transformed by isospin rotation is sufficiently established, see discussion in section \ref{sec:ambig}. 

One could wonder, if it is not better to drop the terms of $\sigma$, $f_0$ and $f_2$
from the current parameterization, 
so currents for both channels would have the same functional form. 
Indeed, that was often the case. For example, 
parametrizations {\bf TAUOLA BaBar} and {\bf TAUOLA RChL 2012} introduced later,
used identical currents for $\pi^-\pi^-\pi^+$ and $\pi^0\pi^0\pi^-$. 
In {\bf TAUOLA CLEO} the same current for both channels was used disregarding 
the fact that it was seemingly introducing $\sigma$ of a charge $\pm$2.
In fact, it was simply assumed that $\sigma$ was just a feature of the current
parametrization and {\it not} a real physical state. 
This is supported by the observation that on Dalitz plot from Fig. 3 of Ref.~\cite{Shibata:2002uv} 
there seems to be a resonance-like structure in $s_3$, the invariant mass of the $\pi^-\pi^-$ pair. 
In corresponding plot, Fig. 5 of Ref.~\cite{Asner:1999kj} for $\pi^0\pi^0\pi^-$ case, 
the same structure can be seen in $\pi^0\pi^0$.
It is most likely originating in part from interferences, 
but it can be parameterized as a resonance\footnote{
See also Section~\ref{sec:example} where we discuss some 
possible origins of unexpected structures.}.
Let us now look into numerical differences between basic variants.
  
\subsection{Numerical differences of {\bf TAUOLA CLEO} and {\bf CLEO publ.}}

\begin{figure}[h!]
\centering
\includegraphics[scale=.650]{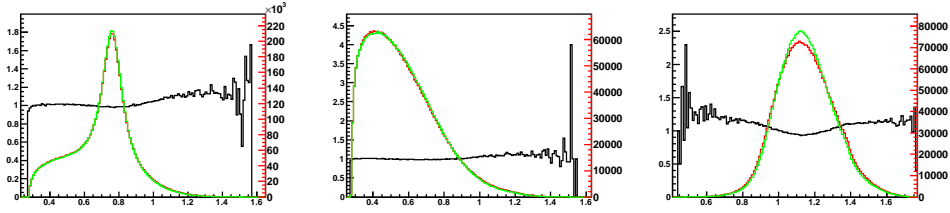}
\caption{Invariant mass, GeV units, of (from the left) $\pi^-\pi^+$, $\pi^-\pi^-$, $\pi^-\pi^-\pi^+$ 
in {\bf TAUOLA CLEO} with numerical constants shifted to values of Ref.~\cite{Asner:1999kj} 
(red) and {\bf TAUOLA CLEO} with default parameterization (green). 
For black line which represent ratio of red plot to green one, left hand side scale should be used. 
Right hand side scale represent number of events per bins of 10 MeV. 
\label{fig:5}}
\end{figure}

The MC sample presented in \cite{Asner:1999kj} is not available now, 
therefore we can only try to recreate results. 
To do so, we need to identify all differences in initializations. 
It appears, that only a few parameters differ, see Table \ref{tab:vsCLEO}.
Resulting difference is small but clear on, see Fig.~\ref{fig:5}.
Note, that these two variants used fixed values for masses and widths 
of resonances (while RChL and BaBar models fitted them). 
We can only speculate it is either due to stability\footnote{
Overpopulation of parameters may affect stability. 
If that was not the case, we may suspect lack of computing power,
as additional parameters increase CPU time needed for fit to converge substantially.
} of fit procedure or
simply the belief that use of results from other (non $\tau$) measurements is better. 
Apart from extreme kinematic regions, 
the largest differences can be observed in $\pi^-\pi^-\pi^+$ distribution.

We can also investigate differences in terms of three-dimensional plots.
Fig.~\ref{fig:CLEOdefault} presents Dalitz plots obtained from {\bf TAUOLA CLEO}. 
As we do not have data nor MC generated sample from the CLEO collaboration, only 
human eye comparison is possible with their published results.
Mentioned Fig.~\ref{fig:CLEOdefault} of our paper should coincide\footnote{
Note, results of Ref.~\cite{Asner:1999kj} measurements are for the $\pi^0\pi^0\pi^-$ 
mode, but the same functional form of hadronic current is used for the $\pi^-\pi^-\pi^+$ in 
{\bf TAUOLA CLEO}, that is why such comparison is valid.} 
with Fig. 5 of. Ref.~\cite{Asner:1999kj}. 
No convincing differences can be seen.

\subsection{Numerical differences of {\bf TAUOLA CLEO} and {\bf TAUOLA CLEO isospin intricate}}\label{sec:numCLEO}

\begin{figure}[h!]
\centering
\includegraphics[scale=.650]{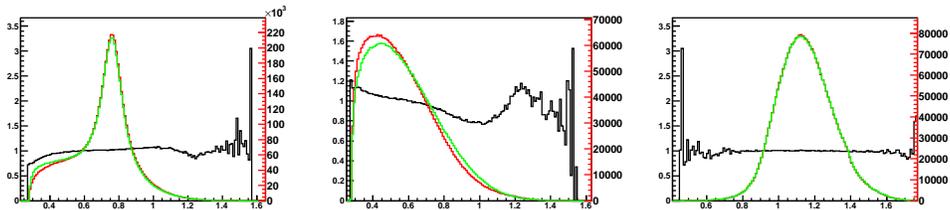}
\caption{Invariant mass of (from the left) $\pi^-\pi^+$, $\pi^-\pi^-$, $\pi^-\pi^-\pi^+$ in {\bf TAUOLA CLEO} 
(red) and {\bf TAUOLA CLEO isospin intricate} (green). 
For details of histograms definition see caption of Fig.~\ref{fig:5}. 
\label{fig:r1vsr0}}
\end{figure}

These two currents have the same values of parameters but differ 
by functional form of hadronic current for $\tau \to \pi^-\pi^-\pi^+\nu_\tau$. 
{\bf TAUOLA CLEO} is represented by equations \eqref{eq:7b}, \eqref{eq:8b},
while {\bf TAUOLA CLEO isospin intricate} by eqs. \eqref{eq:7}, \eqref{eq:8}. 
The differences are large, see Fig.~\ref{fig:r1vsr0}, 
judging with today standards for measurements, but not larger than of relatively 
recent parameterizations {\bf TAUOLA RChL} and {\bf TAUOLA RChl 2012} shown later in Fig.~\ref{fig:massesRChL}. 
In fact they are at the ~10\% level, 
that is at the level of the theoretical uncertainty for use of isospin symmetry \cite{Actis:2010gg}.

\section{Further variants of currents used by CLEO} \label{sec:variants}

Currents described in this paper are not the only ones developed by experimental collaborations. 
We would like to hint why certain solutions were used and 
documented in publications while others seem to be forgotten. 

We should start the discussion with {\bf CLEO publ}. 
This current is fully described in Ref.~\cite{Asner:1999kj}. 
But the very same publication mentions a couple of other different variants, 
stating for some of them even better agreement with the data. 
So why is that any of those did not become main model of the collaboration?  

First, we should mention variant of model where $\sigma$ mass and width were fit parameters. 
Best agreement\footnote{Negative doubled logarithmic likelihood was 43 units below that of nominal fit.} 
with the data was obtained with $M_\sigma=555$ MeV and $\Gamma_\sigma=540$ MeV 
(page 22 of Ref.~\cite{Asner:1999kj}). 
Such fit caused shift in all couplings not exceeding 20$\%$. 
We may again only speculate that it was believed that it is better to use 
theoretically predicted or experimentally well established values. 
Fitting of sigma mass and width was probably performed 
in order to check how seriously this resonance should be treated. 
Result: different values than used for nominal model 
($M_\sigma=860$ MeV and $\Gamma_\sigma=880$ MeV), was most likely not very encouraging. 

Secondly, there is mentioned in Ref.~\cite{Asner:1999kj} a study on pseudo-scalar contribution from $\pi$(1300). 
It also improved the fit result, but with available 
data its existence could not be established with unquestioned significance. 
The very same reasoning was present in study of $a'_1$ contribution, 
which also was not included in nominal model. 

PhD thesis \cite{Hinson:2001cc} presents similar study of pseudo-scalar $\pi$(1300) in 
$\tau \to \pi^-\pi^-\pi^+\nu_\tau$ decay. 
Different ways of inclusion of scalar current were tested there. 
Similarly agreement with the data got better, but was not considered significant enough. 
In this study hadronic current used for non-scalar part was the same as 
{\bf TAUOLA CLEO isospin intricate}. 
It suggests again, that at some point the current without 
pseudo-scalar contribution was used by the collaboration. 

Another thing under consideration by CLEO was the resonance radius. 
Their study showed small importance. 
Therefore, it was set to 0 in order to simplify the model. 
Keeping things as simple as possible seem to be important 
for many choices of collaborations.

The BaBar collaboration had all CLEO currents available, yet decided to use simpler currents. 
Apart from simplicity, one can suspect another reason behind such a choice. 
The BaBar collaboration use commonly one-dimensional histograms in analysis 
even in our times \cite{Nugent:2009zz}. 
Inclusion of any resonances other than $\rho$ and $\rho'$ is not to be easily exploited
without three-dimensional distributions. 
On Fig.~\ref{fig:masses5} one can see, that differences are not very compelling. 
They are mostly below 10$\%$. Comparison of data and theoretical model 
on scatter-grams like Fig.~\ref{fig:ratio2} 
can show more clearly importance of other resonances,  
but it was never performed. On the last figure, ratio 
of Dalitz plots coming from BaBar model and CLEO model is shown. 
While ratio of invariant mass distributions was mostly between 0.9-1.1, 
ratio in some areas of Dalitz plots escapes 0.5-2 range. 


\section{Comparison of {\bf TAUOLA CLEO}, {\bf TAUOLA CLEO isospin intricate} and {\bf TAUOLA BaBar}}\label{sec:vsBaBar}

In this section comparison of CLEO and BaBar model used in {\tt TAUOLA} is shown. 
Parameters are divided into three groups depending on the way
they are introduced into the code and how they were used. 
Tab.~\ref{tab:vsCLEO1} presents those defined globally, outside form factors.
Table \ref{tab:params} presents those introduced inside hadronic current only. 
Parameters in this table were kept constant during the fitting 
procedure of CLEO, but were fitted for BaBar modeling.
Table \ref{tab:fitparams} stores parameters fitted to the data. 
They are also defined inside hadronic current.
Note, that some parameters can be different for diverse
parts of code, for example mass of $\rho$ resonance. 
This may be misleadingly worrisome, 
but it should not be considered as an error. 
Parameters used outside 
hadronic form factors affect optimization 
of efficiency of phase-space generators, 
that is essentially speed of generation only, or are well established constants like $m_\pi$. 
Their variation does {\it not} affect output distributions, or variations are simply unphysical.

\begin{table}[h]
\centering
\begin{tabular}{|c|c|}
\hline
TAUOLA CLEO & TAUOLA BaBar \\ \hline
 & $m_{\pi^0}$ = AMPIZ =0.134996 GeV \\ 
 & $m_{\pi}$ = AMPI =0.139570 GeV \\
 & $M_{\rho}$ = AMRO =0.775 GeV (used in GFUN) \\
 & $M_{a_1}$ = AMA1 =1.251 GeV \\
 & $\Gamma_{a_1}$ = GAMA1 =0.599 GeV \\
$\beta$ = BET =0 & $\beta$ = BET =0\\
\hline
\end{tabular}
\caption{Parameters defined outside form factors (stored in INIMAS). \label{tab:vsCLEO1}}
\end{table}

\begin{table}[h]
\centering
\begin{tabular}{|c|c|}
\hline
TAUOLA CLEO & TAUOLA BaBar \\ \hline
$M_{\rho}$ = MRO =0.7743 GeV & $M_{\rho}$ = ROM =0.773 GeV (used in Breit-Wigner) \\
$\Gamma_{\rho}$ = GRO =0.1491 GeV & $\Gamma_{\rho}$ = ROG =0.145 GeV \\
$M_{\rho'}$ = MRP =1.370 GeV & $M_{\rho'}$ = ROM1 =1.370 GeV \\
$\Gamma_{\rho'}$ = GRP =0.386 GeV & $\Gamma_{\rho'}$ = ROG1 =0.510 GeV \\ 
$M_{f_2}$ = MF2 =1.275 GeV & $\beta_1$ = BETA1 =-0.145 \\
$\Gamma_{f_2}$ = GF2 =0.185 GeV & \\ 
$M_{f_0}$ = MF0 =1.186 GeV & \\
$\Gamma_{f_0}$ = GF0 =0.350 GeV & \\ 
$M_{\sigma}$ = MSG =0.860 GeV &  \\
$\Gamma_{\sigma}$ = GSG =0.880 GeV &  \\
$m_{\pi_0}$ = MPIZ =0.134976 GeV &  \\
$m_{\pi}$ = MPIC =0.139570 GeV &  \\
$m_{K^0}$ = MK =0.496 GeV & \\
$M_{K^*}$ = MKST =0.894 GeV & \\
$M_{a_1}$ = AMA1 =1.275 GeV & \\
$\Gamma_{a_1}$ = GAMA1 =0.700 GeV & \\
$\beta_1$ = BT1 = 1 & \\
\hline
\end{tabular}
\caption{Parameters used in definition of hadronic currents in {\tt TAUOLA}; 
both variants of CLEO parameterization and BaBar parameterization. 
This table stores parameters that were not fitted by the CLEO collaboration, 
but were taken from PDG tables or theoretical predictions. 
Corresponding BaBar parameters are stored here for completeness.\label{tab:params}}
\end{table}

\begin{table}[h]
\centering
\begin{tabular}{|c|}
\hline
$\beta_2$ = BT2 = 0.12$e^{i 0.99 \pi}$  \\
$\beta_3$ = BT3 = 0.37$e^{-i 0.15 \pi}$  \\
$\beta_4$ = BT4 = 0.87$e^{i 0.53 \pi}$  \\
$\beta_5$ = BT5 = 0.71$e^{i 0.56 \pi}$  \\
$\beta_6$ = BT6 = 2.10$e^{i 0.23 \pi}$  \\
$\beta_7$ = BT7 = 0.77$e^{-i 0.54 \pi}$  \\
$C_{3\pi}$ = C3PI = $0.2384^2$  \\
$C_{K^*}$ = CKST = $4.7621^2 C_{3\pi}$  \\
\hline
\end{tabular}
\caption{Parameters used in definition of CLEO hadronic current in 
{\tt TAUOLA} obtained by the CLEO collaboration from fit to the data.
\label{tab:fitparams}}
\end{table}

The {\bf TAUOLA CLEO} code is essentially as described in Ref.~\cite{Asner:1999kj} and named 'nominal fit'. 
It differs by some numerical constants only, see Table \ref{tab:vsCLEO}. 
The CLEO collaboration did not fit masses and widths of resonances but only coupling constants $\beta_i$.
Such choice was driven by theoretical assumptions and measurements of masses elsewhere. 
This may be the best solution when we 
lack good enough measurements of state used in model.
Let us recall, Section~\ref{sec:variants}, 
how insensitive the model can be to the $\sigma$ meson properties.


\subsection{Numerical comparison of {\bf TAUOLA CLEO} and {\bf TAUOLA BaBar}}

\begin{figure}[h!]
\centering
\includegraphics[scale=.650]{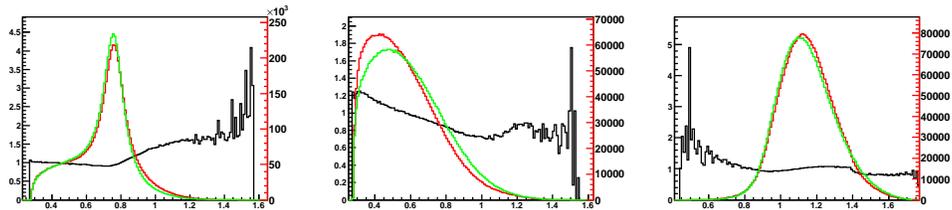}
\caption{Invariant mass of (from the left) $\pi^-\pi^+$, $\pi^-\pi^-$, $\pi^-\pi^-\pi^+$ in {\bf TAUOLA CLEO} (red) and {\bf TAUOLA BaBar} (green). 
For details of histograms definition see caption of Fig.~\ref{fig:5}.
\label{fig:masses5}}
\end{figure}

Here, we present numerical comparison of {\bf TAUOLA CLEO} and {\bf TAUOLA BaBar}.
On Fig.~\ref{fig:masses5}, distributions of invariant masses of 
$\pi^-\pi^+$, $\pi^-\pi^-$, $\pi^-\pi^-\pi^+$ are shown.
Looking at the plots one notice that except the lack of some resonances mentioned earlier, 
also lack of $K^*K$ threshold in BaBar model is visible in distribution of 
$\pi^-\pi^-\pi^+$ mass (and even more visible on ratio plot) in 1.4 GeV region.
Moreover, different $a_1$ mass plays a role. 
In addition, on Fig.~\ref{fig:ratio2} of Appendix \ref{sec:3dplots}, ratio of Dalitz plots is shown. 
Lack of $f_0$, $f_2$, $\sigma$ resonances in model used by BaBar can be also seen there.
Differences are higher than for comparison of {\bf TAUOLA CLEO isospin intricate} and {\bf TAUOLA CLEO},
see Fig.~\ref{fig:5}. Invariant mass of $\pi^-\pi^+$ system exhibits biggest difference in terms of ratio.

\subsection{Numerical comparison of {\bf TAUOLA CLEO isospin intricate} and {\bf TAUOLA BaBar}} 

\begin{figure}[h!]
\centering
\includegraphics[scale=.650]{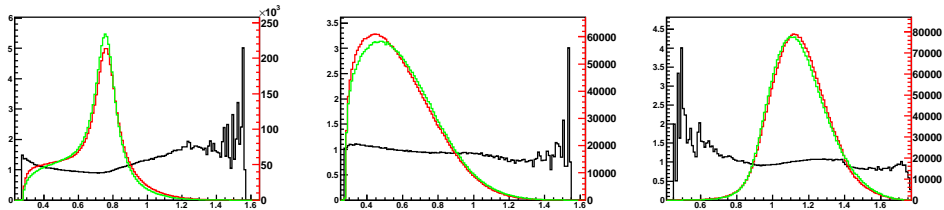}
\caption{Invariant mass of (from the left) $\pi^-\pi^+$, $\pi^-\pi^-$, $\pi^-\pi^-\pi^+$ in {\bf TAUOLA CLEO isospin intricate} (red) and {\bf TAUOLA BaBar} (green). 
For details of histograms definition see caption of Fig.~\ref{fig:5}.
\label{fig:cleor1vsbabar}}
\end{figure}

For completeness we provide comparison of {\bf TAUOLA CLEO isospin intricate} and {\bf TAUOLA BaBar}.
Results are very similar to previous comparison and can be seen on Fig. \ref{fig:cleor1vsbabar}. 
We do not discuss those results as no new aspects appear. 
Largest difference is present in $\pi^-\pi^+$ invariant mass spectrum.

\section{Comparison of {\bf TAUOLA CLEO} and {\bf TAUOLA RChL}}\label{sec:vsRChL}

\begin{figure}[h!]
\centering
\includegraphics[scale=.650]{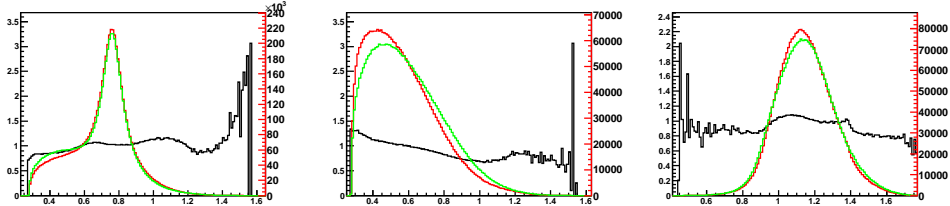}
\caption{Invariant mass of (from the left) $\pi^-\pi^+$, $\pi^-\pi^-$, $\pi^-\pi^-\pi^+$ in {\bf TAUOLA CLEO} (red) and {\bf TAUOLA RChL} (green). 
For details of histograms definition see caption of Fig.~\ref{fig:5}.
\label{fig:masses6}}
\end{figure}

In this section, numerical comparison of CLEO and RChL model used in {\tt TAUOLA} is shown. 
We will not recall analytic form of RChL current. 
Full description of {\bf TAUOLA RChL 2012} current is given in Ref.~\cite{Shekhovtsova:2012ra} 
(see also \cite{Dumm:2009va}), 
and for {\bf TAUOLA RChL} in Ref.~\cite{Nugent:2013hxa}. 
CLEO and RChL models differ a lot in a way of construction and also in parameters, 
but one may easily guess what affects the spectra in a way as seen on Fig.~\ref{fig:masses6}.
Most importantly, inclusion and parameters of $\sigma$ resonance are different. 
The $\sigma$ of lower mass and width than in the CLEO model, 
affects mainly low mass region of $M(\pi^-\pi^+)$ in RChL,
while in CLEO model it affects the whole spectra. 
Second, clearly visible difference appears in $\pi^-\pi^-\pi^+$ invariant mass distribution.
It comes from different modeling of $a_1$. 
CLEO translated theoretical modeling into polynomial of eq. \eqref{eq:a1cleo}.
RChL keeps direct model coded, but parameters of $K^*$ meson were not fitted, 
as the $\pi^-\pi^-\pi^+$ channel should not be very sensitive to them. 
Reasonable modification of those parameters would require simultaneous fit of the $KK\pi$ channel.
Such fit could possibly reduce difference in area of the $K^*K$ threshold mentioned in previous section.

In Fig.~\ref{fig:ratio3} of appendix~\ref{sec:3dplots},  
we can find three-dimensional ratio of RChL to CLEO model results. 
On that plot, we can see areas where difference between the models exceeds factor of 2. 
This is much bigger difference than seen on one-dimensional distributions. 
As CLEO model was fitted to three-dimensional distribution it is 
a very worrisome observation from the RChL perspective. 
It may hint that even though one-dimensional BaBar data distributions are described well~\cite{Nugent:2013hxa}, 
underlying physics model may be imperfect. 
Actual judgment should not be completed without 
comparison with experimental distributions of higher dimensionality.


\subsection{Comparison of {\bf TAUOLA RChL 2012} and {\bf TAUOLA RChL}}\label{sec:RChLcomp}

\begin{figure}[h!]
\centering
\includegraphics[scale=.650]{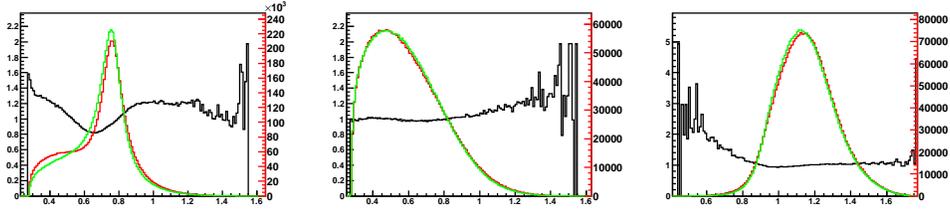}
\caption{Invariant mass of (from the left) $\pi^-\pi^+$, $\pi^-\pi^-$, $\pi^-\pi^-\pi^+$ 
in {\bf TAUOLA RChL} (red) and {\bf TAUOLA RChL 2012} (green).
For details of histograms definition see caption of Fig.~\ref{fig:5}.
\label{fig:massesRChL}}
\end{figure}

For completeness, we present comparison of first version of RChL model and a new one. 
Because of the comparison with the data, 
especially inability to obtain satisfactory agreement in $\pi^-\pi^+$ invariant mass, 
when all invariant mass distributions were fitted simultaneously {\bf TAUOLA RChL 2012} was modified. 
Those two models differ mainly by inclusion of $\sigma$ resonance in the {\bf TAUOLA RChL} version.
Other differences are model parameters and minor correction for Coulomb interaction.
Resulting differences in distributions can be seen on Fig.~\ref{fig:massesRChL}.

At this point we must emphasize that those two versions of RChL model were fitted to different data-sets. 
Invariant mass of $\pi^-\pi^-$ system became available for fits only after
development of {\bf TAUOLA RChL 2012}. 
This model was fitted to invariant masses of $\pi^-\pi^+$, $\pi^-\pi^-\pi^+$ systems only, 
while {\bf TAUOLA RChL} to invariant masses of $\pi^-\pi^+$, $\pi^-\pi^-$, $\pi^-\pi^-\pi^+$. 
It was not the only difference. Also binning of histograms changed. 
New version was fitted to histograms of 10MeV bin width\footnote{
This kind of binning is used in all one-dimensional plots of our paper.}. 
Distributions used in fit of {\bf TAUOLA RChL 2012} had bin width of 20MeV.

\section{Numerical effects of interferences within currents \label{sec:interf}}

So far, we have concentrated on presentation of different currents used 
for Monte Carlo simulations of $\tau$ decays. We have looked into their 
analytical form as well as the numerical values of the parameters 
fixed by confrontations with experimental data. 
Comparisons obtained from different parameterizations were reported as well.
In the present Section let us 
look into numerical consequences of some features of the currents. 
We will concentrate on interferences between distinct parts of the current, 
which can be separated, with the help of isospin transformation.

Final states of long-living scalars with given multiplicity and charge combination,
 can be formed through intermediate resonances used in
 description  of hadronic current. 
One can not think of them as sum of non coherent contributions.
Interferences represent 
important aspect of hadronic current which should be understood and controlled.
Even if used for fits experimental distributions are correctly reproduced,
unnatural response to modifications of individual contributions to 
the current and interference can hint for modeling flaws. 

In principle, addition of a new resonance to the hadronic current should result
in negative interference below the peak and positive above\footnote{
That is true in case when resonances come in with the same sign. 
In our case signs are opposite.}. 
There is no clear prediction on how strong such resonance should interfere.
In case of a narrow intermediate resonance we expect such interference to cancel out 
in contributions to total rates or relatively inclusive distributions. 
Especially in case of wide resonances, we may model a process 
with something that is not of proper dynamical nature but 
still describes data quite well, simply by accident.




In this section, we will investigate RChL and CLEO modeling of the three-pion channels.
We concentrate on interference of part depending on isoscalars\footnote{
Contribution only from $\sigma$ in RChL, while for CLEO from $\sigma, f_2, f_0$.} 
with the remaining part of the current. 
These two parts of hadronic current transform differently under isospin rotation.
In principle it is also a test on how justified (or not) is isospin rotation 
from $\pi^0\pi^0\pi^-$ to $\pi^-\pi^-\pi^+$ currents.

Let us look at the contribution to the rate. 
For $\pi^0\pi^0\pi^-$, the $\rho$ part, 
the $\sigma, f_2, f_0$ part and their interference contribution are respectively\footnote{
When the same current is used for $\pi^-\pi^-\pi^+$ corresponding contributions are 75.3\%, 4.3\% and 20.4\%.
} 75.1\%, 4.4\% and 20.5\%.
Usage of {\bf TAUOLA CLEO isospin intricate} current for $\pi^-\pi^-\pi^+$ 
gives 74\%, 5.8\% and 20.2\%. Interference looks slightly different for fully differential distributions. 
Then, it is 22.6\% in {\bf TAUOLA CLEO}, and 20.7\% for {\bf TAUOLA CLEO isospin intricate} case, 
when method as explained in Section~\ref{sec:datarep} is used.
For the three-dimensional scatter-grams of the CLEO collaboration style, 
see Appendix \ref{sec:3dplots} for details,
the interference is respectively 22.3\% and 20.6\%. 
This demonstrates nearly perfect sensitivity of such a method. 

\begin{figure}[h!]
\centering
\includegraphics[scale=.650]{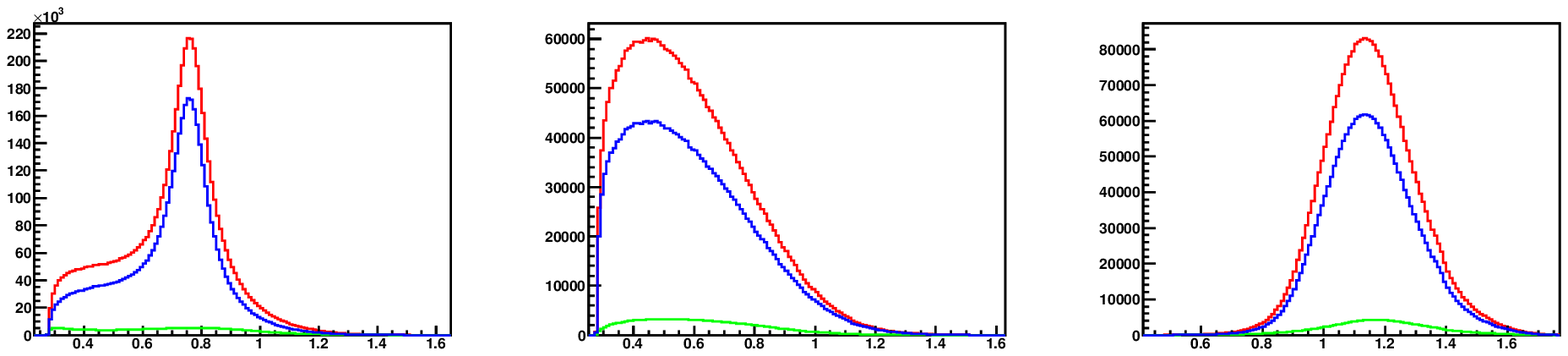}
\includegraphics[scale=.650]{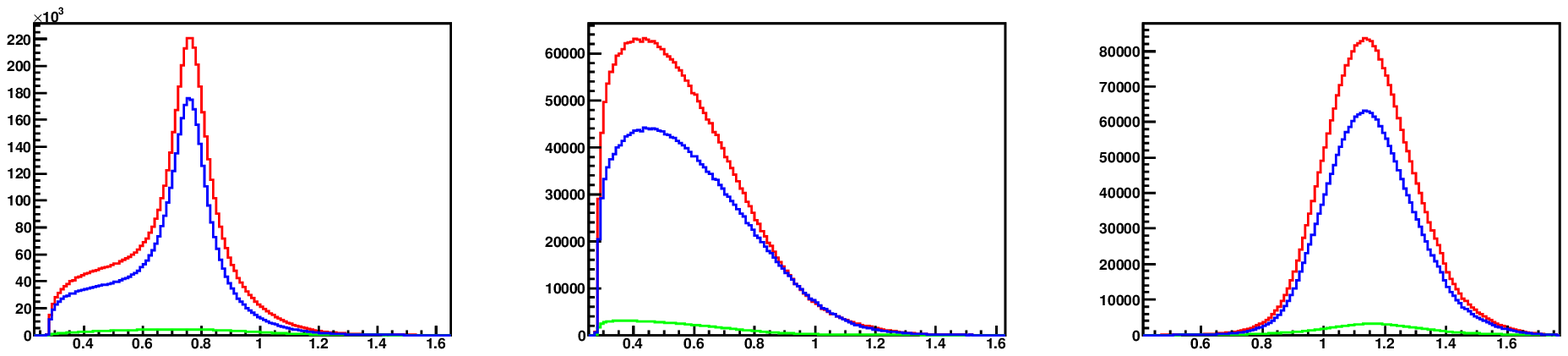}
\caption{Invariant mass of (from the left) $\pi^-\pi^+$, $\pi^-\pi^-$, $\pi^-\pi^-\pi^+$ 
using {\bf TAUOLA CLEO isospin intricate} (top) and {\bf TAUOLA CLEO} (bottom). 
Green plot represents contribution from isoscalars ($\sigma, f_2, f_0$) 
and blue from remaining part.
Red plot represents the whole current. 
Blue and green plots would sum up to red one, in the absence of interference. 
Percentages of the three contributions are given in Table \ref{tab:interf}.
\label{fig:massint1}}
\end{figure}

\begin{table}[h]
{\tiny
\centering
\begin{tabular}{|c||c|c|c|c|}
\hline 
Current & Absolute width & 1D distributions & 
3D distributions & Absolute width differential \\ \hline 
$\pi^0\pi^0\pi^-$ {\bf CLEO} & 75.1\%  4.4\% 20.5\% & 75.1\%  4.4\% 20.8\% & 75.1\%  4.4\% 22.4\%  & 75.1\%  4.4\% 22.7\%  \\ \hline
$\pi^-\pi^-\pi^+$ {\bf CLEO} & 75.3\%  4.3\% 20.4\% & 75.3\%  4.3\% 20.7\% & 75.3\%  4.3\% 22.3\% & 75.3\%  4.3\% 22.6\%  \\ \hline
$\pi^-\pi^-\pi^+$ {\bf CLEO isospin intricate} 
& 74\%  5.8\% 20.2\% & 74\%  5.8\% 20.2\% & 74\%  5.8\% 20.6\% & 74\%  5.8\% 20.7\% \\ \hline
$\pi^-\pi^-\pi^+$ {\bf RChL} & 97.5\%  6.5\% 4\% & 97.5\%  6.5\% 11\% & 97.5\%  6.5\% 18.2\% & 97.5\%  6.5\% 18.5\%\\ \hline
$\pi^0\pi^0\pi^-$ {\bf RChL} & 94\% 1.3\% 4.7\% & 94\% 1.3\% 7.4\% & 94\% 1.3\% 9.5\% & 94\% 1.3\% 10\% \\ \hline
\end{tabular} }
\caption{Comparison of contributions from the isoscalars from the
remaining part of the current and from interference,  
depending on a way of investigation. 
In each cell, first number is from the  
spin one intermediate states, 
second is from isoscalars and the third one 
is interference. For 1-D (one-dimensional) distributions 
interference effect of the histogram with largest effect is taken. 
In the case of "Absolute width differential", 
average of (module of) interference over all events is shown. 
\label{tab:interf}}
\end{table} 

In Table~\ref{tab:interf} we collect such results also for other versions of the currents.
One must remember that the interference visible in histograms is somewhat dependent on binning. 
We do not have the same binning as the CLEO collaboration; 
they had less bins per energy unit in every distributions. 
Therefore, while they probably could have seen the same behavior, 
the interference observed was most likely smaller. 
Interference is an important numerical property of the distributions which
can not be ignored while making the choices for phenomenological 
approach\footnote{ Later, in particular
in Section~\ref{sec:example} we  discuss how resonances are deformed and 
even can produce artificial peaks. 
Therefore, resonances  are not always needed to emulate peaks
present in experimental distribution. 
Also, because shapes of resulting modifications to distributions are 
distinct with respect to, e.g., Gaussian peaks of resonance parameterizations, 
we can be concerned about choice between {\bf TAUOLA CLEO} and {\bf TAUOLA CLEO isospin intricate}. 
This especially, as dominant contribution of isoscalars is a result of interference.
Confrontation with the experimental data should be used to resolve this matter. 
}.

\begin{figure}[h!]
\centering
\includegraphics[scale=.650]{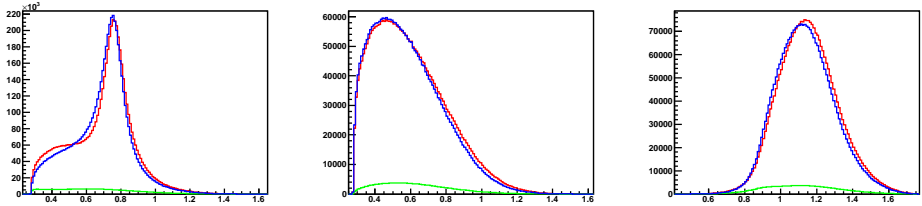}
\caption{Invariant mass of (from the left) $\pi^-\pi^+$, $\pi^-\pi^-$, $\pi^-\pi^-\pi^+$ 
using {\bf TAUOLA RChL} current. 
Green plot represents contribution from isoscalars and blue one from remaining part.
Red plot represents whole current. Blue and green plots should sum up to red one in the absence of interference.
One can see negative interference for RChL model whereas
it was almost everywhere positive for {\bf TAUOLA CLEO isospin intricate}, see Fig.~\ref{fig:massint1}. 
and Table \ref{tab:interf}.
\label{fig:massint2}}
\end{figure}

 
Having discussed CLEO modeling let us investigate the same problem in RChL model.
On Fig.~\ref{fig:massint2}, one can compare contributions from different parts of current 
on one-dimensional distributions. 
We can see that plot follows theoretical expectation on how interference should look. 
Also, there is much less interference than for CLEO models, see Table \ref{tab:interf}. 
Moreover, size of interference for the one-prong and three-prong 
channels is not as similar as in CLEO modeling. 
While both CLEO models have almost the same level of interference in both channels, 
RChL has twice more in the three-prong channel. 
Also, size of $\sigma$ contribution differ a lot between the one- and the three-prong channels, 
while it is close to the same in both CLEO modelings.
Another property worth noting is limited sensitivity of RChL model to interference
when only one-dimensional distributions are used. 
It means that the most recent fit of this model does not fully control\footnote{
The reason why interference effect is better visible in one-dimensional 
distributions in case of CLEO model than in RChL is technically quite 
simple. In the first case the mass of the $\sigma$ is larger so the region of 
phase-space above the peak is restricted, no cancellation occurs.
Nonetheless it is important to realize such features while evaluating 
suitability of experimental input for the model fit. It should not be 
assumed that the conclusion will remain the same for all range of the fit parameters.
}
 influence of $\sigma$ and fit to three-dimensional distributions is strongly desired.

 As both models aim at describing the same decay channels we should be concerned that interplay of 
internal parts of the current is quite different. 
It proves that we do not have good understanding of physics behind it, especially what exactly is 
the $\sigma$ state, in at least one of the two approaches. Both models find it 
important \cite{Asner:1999kj} \cite{Nugent:2013hxa} 
to include $\sigma$, but its description is far from consistent\footnote{
We may speculate that in part, this difficulty  in control contribution from
 isoscalars was a a reason behind introduced later much simpler current in 
BaBar Monte Carlo simulations. Note, that in Ref.~\cite{Nugent:2013hxa} 
parameters of $\sigma$ contributions were featuring errors from the fit larger than for other resonances.
We will return to this point in Section \ref{sec:Properties}.
}.

\section{Interplay of experimental and theoretical input}\label{sec:Properties}

In the process of creating a model for hadronic current to be used in MC there are two aspects.
Theoreticians provide model for hadronic currents and experimentalists use it and fit parameters. 
Usually, theoreticians concentrate on consistency of  
formulation and prefer to avoid ad hoc phenomenological corrections. 
Very often, model building is disconnected from data analysis. 
On the other hand, experimentalists would like to have model perfectly describing the data. 
At the same time, they must have control over any possible sources of experimental errors. 
Therefore, they may prefer to sacrifice some model assumptions 
for the sake of simplicity and more straightforward systematics analysis\footnote{
In particular that there are no strongly correlated parameters in the fits.}. 
As experiment controls data sample it is experimentalists to give 
"stamp of approval" for the hadronic current.  
Such thing was done for $\pi^0 \pi^0 \pi^- $ current of CLEO modeling, 
but is lacking for $\pi^- \pi^- \pi^+ $ case. 
Only conference contributions describe some modelings of this channel. 
In case of the CLEO modeling we may only speculate if there were some unresolved problems 
with this channel or it was simply lack of the manpower that stopped the collaboration publication from appearing.
In the following subsections we will discuss some aspects of model-building that can play a role.

\subsection{Ambiguities of experimental inputs}\label{sec:ambig}

Isospin symmetry between $\tau$ decay channels,
can be restricted to a relation of final states only. 
As a consequence, for intermediate states,  
such a choice could be understood as including contribution of e.g., 
$\sigma \to \pi^- \pi^-$ for the channel 
$\tau \to \nu_\tau \pi^- \pi^- \pi^+ $.
This of course would mean an error. 
However, as $\sigma$ was not well established (and very broad) 
resonance at the time of the CLEO work, this may not be unreasonable. 
Back then,  $f_0$(400-1200) \cite{Caso:1998tx} with width ranging from 400-1000 MeV was expected. 
Later on, in years 2002-2012 $f_0$(600) was expected \cite{Hagiwara:2002fs}. 
Recently, in 2012 it was shifted to $f_0$(500) \cite{Beringer:1900zz}, 
thanks to numerous calculations, e.g., \cite{GarciaMartin:2011jx, Caprini:2005zr} 
claiming both mass and width close to 500 MeV. 
In current measurements this resonance remain extremely wide, while the CLEO 
collaboration used even wider (880 MeV) and more massive (860 MeV) 
$\sigma$, as of Ref.~\cite{Tornqvist:1995kr}.

That is why, from the experimentalist point of view, 
one can think of this amplitude contribution as a heuristic 
modification for the current, of not much quantum number meaning\footnote{
In Ref.~\cite{Asner:1999kj} we read: {\it The form factors as defined in Eqn. 
A2 (Equation A2 of Ref.~\cite{Asner:1999kj} corresponds to our eq. \eqref{eq:Hadcur}) 
do not have simple correspondence with those that can be associated 
with specific resonant contributions to the hadronic current}, which seem to support our statement.}. 
Such interpretation may hold, unless it is excluded by comparisons with experimental data. 
In the meantime, such an option in the Monte Carlo, 
can be helpful to study if the contribution is indeed the $\sigma$. 
Alternatively, it is possible to think that broad $\sigma$ is just a pretext to correct current of 
$a_1$, $\rho$ intermediate states, thus may be identical for one- and three-prong $\tau$ decay modes. 
Similarly one can think of $f_0$. The $f_2$ is not as broad, 
but even in this case contributions to $s_1$, $s_2$ spectra are not clearly localized in the phase-space. 
We should also mention that both $f_2$ and $f_0$ have limited phase-space for decay, 
therefore their Breit-Wigner distributions are strongly deformed. 
On top of that, in Section~\ref{sec:interf} we have shown that isoscalars 
in the models we have presented, do not 
contribute as separate resonances to the decay. 
They contribute positive interference over almost whole spectra.
That is why, such imposed isospin symmetry understood as use of the same current for the
$\pi^- \pi^- \pi^+ $ and the $\pi^0 \pi^0 \pi^- $ decay channels of $\tau$ can be justified. 
Let us recall that isospin symmetry may be broken and of 
limited predictive power starting from 5-10\% 
precision level (Ref.~\cite{Actis:2010gg} Section 5.11). 
Also, there is very small contribution to the currents from isospin violating decays\footnote{
In the $\tau\to 3\pi \nu_\tau $ decay mode isospin violating signals are expected \cite{Mirkes:1997ea}.
There are two decays that can feed $3\pi$ spectrum.
For the three-prong channel it is $\tau\to \omega\pi^-\nu_\tau$ 
with subsequent $\omega \to \pi^-\pi^+$ decay. 
The one-prong channel can be fed by the decay $\tau\to \eta\pi^0\pi^-\nu_\tau$ (due to $\eta-\pi^0$ mixing).
Those decays are estimated to give small contribution 
to the total rate at the level of 0.4\% and $10^{-3}\%$, respectively. 
Studies of $\tau\to \eta\pi^0\pi^-\nu_\tau$ case were also done within RChL scheme \cite{Dumm:2012vb}.}.


\subsection{Further concerns  for model-data confrontation}\label{sec:example}

\begin{figure}[h!]
\centering
\includegraphics[scale=.450]{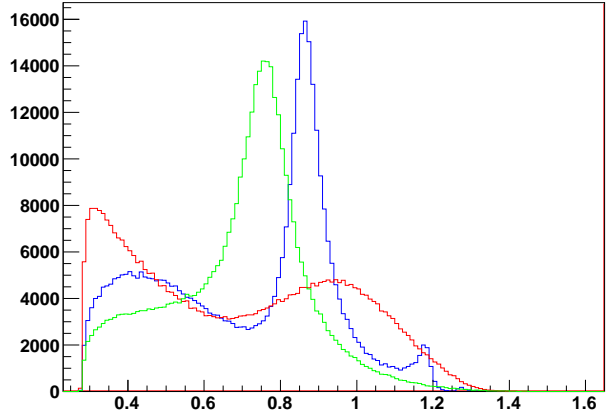}
\caption{Invariant mass of $\pi^-\pi^+$ system in $\tau \to \pi^-\pi^-\pi^+$ 
decay using {\bf TAUOLA CLEO isospin intricate} (green), 
current with multiplied 10 times amplitude from isoscalars (red) and 
also with width of isoscalars reduced 10 times (blue).
\label{fig:example}}
\end{figure}

Let us look into details of the  {\bf TAUOLA CLEO isospin intricate} current, 
and the decay channel $\tau \to \pi^-\pi^-\pi^+\nu_\tau$ . 
The contribution from isoscalars may be missed from the
distribution of $\pi^-\pi^+$ invariant mass spectrum: 
no clean features are seen, Fig.~\ref{fig:example} green line. 
If we artificially increase by a factor of 10 isoscalars amplitude, 
they dominate the shape of $\pi^-\pi^+$ invariant mass spectrum, 
see red line, but clear peaks are still absent.
Position of the peaks is shifted with respect to
the masses used in the propagator\footnote{See formulae \eqref{eq:8b}, \eqref{eq:7b}, 
and Tables \ref{tab:vsCLEO1}, \ref{tab:params}, \ref{tab:fitparams}
for the definition of $\rho$, $\sigma$, $f_2$, and $f_0$ propagators as used by the CLEO collaboration.}
of hadronic current. One can see, blue line, that two peaks emerge clearly 
if widths of isoscalars are reduced by a factor of 10 as well. 
Looking at all three lines, one realizes that appearing low energy peak/deformation present 
in {\bf TAUOLA CLEO isospin intricate} (in other modelings as well) is in a big part indirect. 
There are two possible combinations of $\pi^-\pi^+$ and both are put into the histogram.
Lower peak-like bump comes from small phase-space available for $\pi^-$ when system 
of other $\pi^-$ and $\pi^+$ is produced with high invariant mass. 
That forces low invariant mass bump for the $\pi^-\pi^+$ 
system, even though those two pions do not originate from a common intermediate resonance. 

Resonances at the end of the phase-space represent challenging part of the 
description. As we could see from Fig.~\ref{fig:example}, their shapes are substantially
deformed from the ones of Breit-Wigners. 
The other effect is also possible. 
In Ref.~\cite{Ishikawa:1997ma} it was demonstrated how deformations 
due to resonances acting as phase-space constraints, may result even in artificial 
resonance-like peaks. This observation is another reason why one may be 
hesitant to exploit isospin symmetries of currents in an uncritical way.

Another concern is the size of contribution due to interference. 
It may be much larger than the effect of extra resonances alone.
At the same time, it is not fully explored by inclusive distributions.
See Table~\ref{tab:interf}; for RChL, interference effect is much larger for the three-dimensional 
histograms than for the one-dimensional ones. 
For the CLEO case, the difference was rather minor.
One has to remember of such effects, 
while introducing new resonances into the currents. 
It must be carefully studied if they affect shapes of the distributions in a controlled way.

Finally, if numerical effects are small and weakly depend on model parameters,
Monte Carlo methods may not be best suited for the studies. 
Especially if systematic errors are to be taken into account \cite{KUKDM}.
If available, semi-analytical methods are better. Therefore, 
we have prepared special arrangement which is briefly presented in Appendix~\ref{sec:analytical}.

\section{Summary}\label{sec:summary}

In this paper, we have investigated several variants of parameterizations of hadronic currents 
used in Monte Carlo programs for simulations of the $\tau$ lepton decays. 
As a reference for comparisons we have used {\bf TAUOLA CLEO} parameterization, 
because of its well established status; 
it is supported by the collaboration publication.
The first investigated variant was {\bf TAUOLA CLEO isospin intricate}, 
as prepared for {\tt TAUOLA} but not commonly used. 
This parameterization is documented by conference contribution only. 
We have evaluated the numerical size of the differences, 
which were found to be of the order of systematic errors 
expected from the use of isospin symmetry alone.
We have looked for the origin of numerical differences. 
We studied both analytic forms of the distributions as well as dependence on numerical constants.

Later, we have presented comparisons with further parameterizations, 
in particular the variants of {\bf TAUOLA RChL 2012}, {\bf TAUOLA RChL} and {\bf TAUOLA BaBar}. 
All distributions were obtained with the help of Monte Carlo generations. 
We have prepared semi-analytical distributions to calculate such results as well. 
This is a necessary step toward fitting with the help 
of multidimensional distributions, using variants of 
model independent methods presented in Ref.~\cite{Mirkes}, 
but adopted to the case, when $\tau$ leptons are relativistic 
(that is when $\nu_\tau$ momentum cannot be reconstructed easily). 

We have discussed limitations and constraints on use of models and symmetries like isospin symmetry 
resulting from insufficiently differential distributions of experimental data. 
We have pointed to large numerical consequences of interferences. 
We have also pointed out problems related to phase-space constraints for decay products of broad resonances.

We should accept that predictions of the 
RChL models may agree with the data with precision not worse than 
$\frac{1}{N_C}$ that is about 30\% (in respect to total rate) only. Indeed, all of the presented 
parameterizations pass such condition and agree with each other at that level as well. 
Precision of the experimental data is at least one order of magnitude better. 
That is why, one has to concentrate the effort on arrangements for convenient 
comparisons of the data with  model predictions. It is important that 
models are prepared in sufficiently flexible way 
(the necessary future adaptations can be introduced fast and correlations of parameters are 
understood) before the fits to experimental data are performed.
Also experimental distributions need to be investigated, 
if they are detailed enough to constrain the underlying physics.

We have discussed consequences of limited dimensionality 
of distributions used in data representations for results of fits.
We could see in Ref.~\cite{Nugent:2013hxa}, 
that the model of $\tau\to 3\pi \nu_\tau $  decay modes, 
prepared only a year earlier \cite{Shekhovtsova:2012ra}
and fitted to two, one-dimensional, mass histograms, turned out 
to predict results for the invariant mas of $\pi^- \pi^-$ pair rather poorly. 
It had to be updated. 
We  are expecting this to repeat once three-dimensional experimental histograms are addressed. 
There is a factor of two discrepancy between RChL models
and CLEO models results for some bins of the three-dimensional distributions 
(which were used by CLEO also as input to their fits). 
One-dimensional histograms may thus not be enough as experimental input for 
precise parameterization of hadronic currents. 
The three-dimensional ones of the type as used by CLEO 
\cite{Asner:1999kj} may be sufficient for  $\tau\to 3\pi \nu_\tau $. 
However,
this conjecture may need to be checked before serious work on the $\tau$ decay channels into 3 scalars 
with kaons, like that of Refs.~\cite{Shekhovtsova:2014hda, Dumm:2009kj} is completed. 
Then, full complexity of hadronic currents is expected, and analysis similar to 
the one of Ref.~\cite{Mirkes} may be indispensable. 
Nonetheless, attempts to construct currents on the basis 
of fits to one-dimensional (three-dimensional CLEO style) distributions are of interest. 
It can demonstrate predictive power of models such as RChL 
beyond $\frac{1}{N_C}$ precision level.
On a practical side, once differences and options are understood 
in context of one-dimensional distributions,
we are better prepared to fit the high precision data of today. 

If operators like those in Ref.~\cite{Mirkes} are needed, then they have 
to be adopted for conditions of $\tau$ leptons produced relativistically 
(making reconstruction of $\nu_\tau$ momentum non trivial). 
Our paper documents also a step of work into such direction,
see also web page \cite{Link:JZfit}.  

Even though the purpose of our paper is to review available options for
hadronic currents used in simulation of $\tau$ decays into 3$\pi$,
some speculations, why such choices were introduced in the past, were unavoidable. 
This is by far incomplete aspect of this work, and may be even misleading
at some points. There is little documentation on internal 
discussions of experimental groups behind actual choice of analytical form
of current parameterizations used in data analysis and publications.
Often complicated and sometimes unfinished activities, where
questions of of great importance on systematic errors 
for multidimensional distributions were risen,
could not be referred. 

In general, we think experiments should be responsible 
for fits approving models quality. We hope our paper will  
contribute to discussion how relations between model developers, experimental
physicists and people working on Monte Carlo and fitting programs should look.

\vskip 3 mm
\begin{center}{\Large  Acknowledgments}\end{center}
\vskip 3 mm

We are indebted to co-authors of our publications and experimental partners as well. 
We would like to thank: Simon Eidelman, Swagato Banerjee, Pablo Roig, Marcin Chrzaszcz, 
Tomasz Przedzi\'nski for discussions and for reading the manuscript.

\providecommand{\href}[2]{#2}
\appendix

\section{Analytical distributions}\label{sec:analytical}

From the reference \cite{Nugent:2013hxa} documenting RChL current and 
its fit to the experimental data it became obvious that having semi-analytical 
distributions available, simplifies and speeds up the fitting procedure. 
As a part of this work we have prepared a program to 
calculate distributions from given hadronic current by integrating the 
structure function $W_A$ (as defined in Ref.~\cite{Mirkes}) within limits of a chosen bin. 
Program uses {\tt IntegralMultiple} function of {\tt ROOT} \cite{Antcheva20092499} libraries allowing 
acquisition of both one-dimensional and Dalitz distributions. 
We have checked the code to give proper results for {\bf TAUOLA CLEO}, 
{\bf TAUOLA CLEO isospin intricate}, {\tt TAUOLA BaBar}, and {\bf TAUOLA RChL}. 
Program offers easy way for re-fit of these models to the data of a form of 
three-dimensional distributions like in CLEO publication. 
Also, it is much easier, than if MC method is used, to perform analysis of statistical and systematical 
errors with the semi-analytical results free of statistical errors themselves. 
See Ref.~\cite{Nugent:2013hxa} for details.

{\tt IntegralMultiple} routine of {\tt ROOT} library has several input parameters. 
In particular, minimum and maximum number of function evaluation requested for each bin. 
Those parameters affect precision and time of integration, 
therefore one has to find a balance between the two. 
We have found that integration has some stability 
problems if too low number of requested points is chosen.
Taking into account time, precision, and stability of integration we choose 
minimum 5000 points and maximum 50 000 points per bin. 
With such setup it takes 6 sec (of CPU time on a typical PC processor) 
for {\bf TAUOLA CLEO} and 16 sec for 
{\bf TAUOLA RChL} to get one-dimensional distributions and 
respectively, 47 and 120 sec to get 
Dalitz plots with the same binning as presented in Appendix~\ref{sec:3dplots}. 
One has to add 2 sec for CLEO and 30 sec for RChL models for reinitialization 
whenever model parameters are changed. If higher precision is needed, 
doubling number of points per bin, requires 50$\%$ more time for three-dimensional plots
while it does not noticeably affect calculation of one-dimensional distributions. 
We expect that CPU time for Dalitz plots calculation can still be optimized. 
Fit with Dalitz plots is more desired, it gives better test of investigated 
model and has simpler systematics calculation; 
each individual event enters only one bin in contrary to  
one-dimensional histograms where it enters each\footnote{
Each event enters histograms for invariant mass of 
$\pi^-\pi^-$, $\pi^-\pi^-\pi^+$ and twice for $\pi^-\pi^+$ systems 
(as there are two possible pairs of $\pi^-\pi^+$).} of four histograms of invariant masses.
In Ref.~\cite{KUKDM} optimization of fitting procedure for 
{\bf TAUOLA RChL} in case of one-dimensional histograms is presented. 
We hope that with similar approach we could prepare framework allowing 
fast fitting to three-dimensional distributions for any theoretical model as well. 
Note, that current initialization of RChL is prepared for multiple channels 
of $\tau$ decays, including not yet published ones with kaons 
\cite{Shekhovtsova:2014hda, Dumm:2009kj}. 
These, were not confronted with the data yet. 
For fitting to $\pi^-\pi^-\pi^+$ data, we can skip running routines needed for these other channels, 
to speed it up to the level which was used in Ref.~\cite{KUKDM}. 
We have not checked if {\tt IntegralMultiple} 
will provide smooth distributions with model parameters variation. 
It is important for fitting algorithms that derivatives 
with respect to model parameters are continuous, 
as we have found while working for Ref.~\cite{Nugent:2013hxa}.

\section{Comparison of {\bf TAUOLA CLEO} and {\bf Pythia CLEO}}\label{sec:vsPythia}

In this appendix, we present comparison of alternative 
implementation of CLEO model for $\tau \to \pi^-\pi^-\pi^+\nu_\tau$ 
as done in {\tt PYTHIA} 8.201, with {\tt TAUOLA} implementation. 
Thanks to comparison we can identify source of differences and estimate its impact on generated event samples. 

By investigation of {\tt Pythia} code, several differences in 
parameterization were found and stored in Table~\ref{tab:vsPythia}.
Parameters not mentioned in that table are the same as for TAUOLA.
Hadronic current of {\bf Pythia CLEO} is of the same form as 
described in Section \ref{sec:CLEO} for {\bf TAUOLA CLEO isospin intricate}. 
Internal arrangements of the code are slightly different. 
Due to this; in structure of the code some constants are folded with others.
Parameter $\Gamma_{a_1}$ is one of such. 
It is not coded explicitly, but we can extract its value: 0.784468371 GeV.
 
\begin{table}[h]
\centering
\begin{tabular}{|c|c|}
\hline
TAUOLA CLEO & Pythia CLEO \\ \hline 
$M_{\tau}$ = 1.777 GeV & $M_{\tau}$ = 1.77699 GeV \\
$M_{a_1}$ = 1.275 GeV & $M_{a_1}$ = 1.331 GeV \\
$\Gamma_{a_1}$ = 0.700 GeV & is calculated, see section text\\
C3PI= $0.2384^2$ & C3PI = $0.2384^2$/1.0252088 \\ \hline
$\beta_3$ = BT3 = 0.37$e^{-i 0.15 \pi}$ & $\beta_3$ = BT3 = $3.7\cdot 10^{-7}e^{-i 0.15 \pi}$ \\
$\beta_4$ = BT4 = 0.87$e^{i 0.53 \pi}$ & $\beta_4$ = BT4 = $8.7\cdot 10^{-7}e^{i 0.53 \pi}$ \\
$\beta_5$ = BT5 = 0.71$e^{i 0.56 \pi}$ & $\beta_5$ = BT5 = $7.1\cdot 10^{-7}e^{i 0.56 \pi}$ \\
\hline
\end{tabular}
\caption{Differences in numerical values of parameters used in {\bf TAUOLA CLEO} and {\bf Pythia CLEO}. 
\label{tab:vsPythia}} 
\end{table}

\begin{figure}[h!]

\centering
\includegraphics[scale=.650]{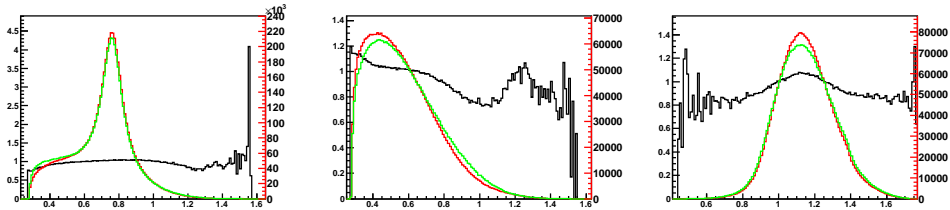}
\caption{Invariant mass of $\pi^-\pi^+$, $\pi^-\pi^-$, $\pi^-\pi^-\pi^+$ in {\bf TAUOLA CLEO} (red) 
and {\bf Pythia CLEO} (green). For details of histograms definition see caption of Fig.~\ref{fig:5}.
\label{fig:masses1}}
\end{figure}

\begin{figure}[h!]
\centering
\includegraphics[scale=.650]{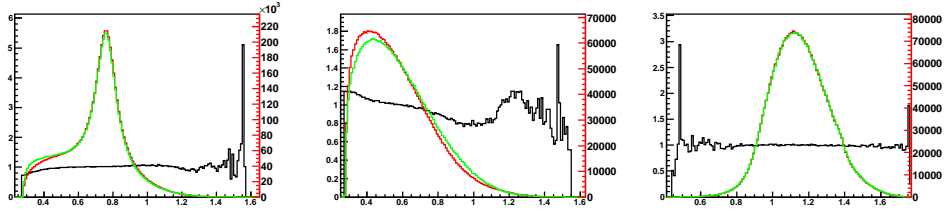}
\caption{Invariant mass of $\pi^-\pi^+$, $\pi^-\pi^-$, $\pi^-\pi^-\pi^+$ in {\bf TAUOLA CLEO} (red) 
and {\bf Pythia CLEO} (green), when the same parameters are enforced for both generators. 
For details of histograms definition see caption of Fig.~\ref{fig:5}.
\label{fig:masses4}}
\end{figure}

Figures \ref{fig:CLEOdefault} and \ref{fig:Pythiadefault} of Appendix \ref{sec:3dplots}
represent Dalitz plots from {\tt TAUOLA} and {\tt Pythia}, respectively. 
Fig.~\ref{fig:ratio1}, 
shows ratio of Fig.~\ref{fig:Pythiadefault} to Fig.~\ref{fig:CLEOdefault}. 
One-dimensional distributions of invariant mass of $\pi^-\pi^+$, $\pi^-\pi^-$, $\pi^-\pi^-\pi^+$ 
for both generators can be seen on Fig.~\ref{fig:masses1}.
There is systematic difference between generators. 
Disagreement may have appeared due to different form of hadronic current, 
or different numerical parameters. 
To test this, for both generators constants of Ref.~\cite{Asner:1999kj} were taken. 
Agreement was not improved substantially, with the exception of invariant mass $\pi^-\pi^-\pi^+$, 
where influence of $a_1$ mass and width turned out to be the main source of difference, 
as can be seen from Fig.~\ref{fig:masses4}. 
In general, agreement only to about 10 \% can be concluded except some less populated bins 
at ends of $\pi^-\pi^+$ mass spectra.

\begin{figure}[h!]
\centering
\includegraphics[scale=.650]{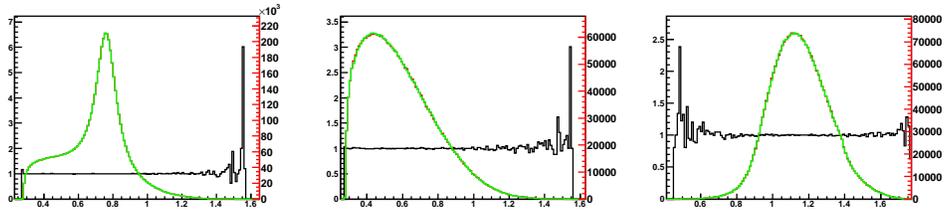}
\caption{Invariant mass of $\pi^-\pi^+$, $\pi^-\pi^-$, $\pi^-\pi^-\pi^+$ in {\bf TAUOLA CLEO isospin intricate}  (red) 
and {\bf Pythia CLEO} (green), when the same parameters are enforced for both generators. 
For details of histograms definition see caption of Fig.~\ref{fig:5}.
\label{fig:masses2}}
\end{figure}

It suggests that functional form of hadronic current is of higher numerical importance, 
than model constants.
To test this, we have used {\bf TAUOLA CLEO isospin intricate} current
while having the same parameters in both generators. 
As in previous test, constants were set that of \cite{Asner:1999kj}. 
Resulting distributions of invariant masses, 
as well as their ratios can be seen in Fig.~\ref{fig:masses2}: 
nearly perfect agreement can be concluded\footnote{
This is also true for Dalitz plots, see Fig.~\ref{fig:ratio1a}.}. 

\begin{figure}[h!]
\centering
\includegraphics[scale=.650]{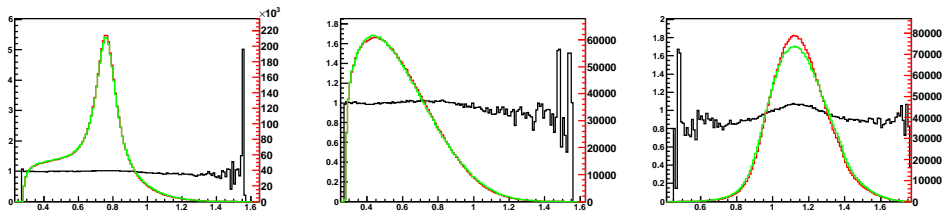}
\caption{Invariant mass of $\pi^-\pi^+$, $\pi^-\pi^-$, $\pi^-\pi^-\pi^+$ in {\bf TAUOLA CLEO isospin intricate} (red) and {\bf Pythia CLEO} (green). 
For details of histograms definition see caption of Fig.~\ref{fig:5}.
\label{fig:masses3}}
\end{figure}

To complete our investigation, comparison of {\bf TAUOLA CLEO isospin intricate} 
and {\bf Pythia CLEO}, with default initializations was performed.
Generators use hadronic current of eqs. \eqref{eq:7}, \eqref{eq:8}, 
and only numerical constants differ as those collected in Table~\ref{tab:vsPythia}. 
Figure \ref{fig:masses3} shows resulting distributions of invariant masses and their ratios. 
One can see the biggest difference in invariant mass of 
$\pi^-\pi^-\pi^+$ due to distinct $a_1$ mass and width, which are of highest importance 
among all numerical constants.

Further, technical checks were performed: the same 
numerical constants and the same but simplified 
hadronic current were installed in both generators. 
Those included: 
\begin{itemize}
\item reducing hadronic current to $F_1$ of eq. \eqref{eq:7b} only ($F_2, F_3=0$).
\item setting hadronic current to constant.
\item reducing hadronic current to Breit-Wigner of $a_1$ only.
\end{itemize}
In all these cases no discrepancies between results 
of {\tt TAUOLA} and {\tt Pythia} generators beyond statistical fluctuations 
were found. Samples of 3 million events were used. 

All considered, differences in produced by {\tt TAUOLA} and {\tt Pythia} distributions for the
$\tau \to \pi^-\pi^-\pi^+\nu_\tau$ decay channel
result only from parameters if the same variant of current is used in {\tt TAUOLA}, 
otherwise generators perform identically. 
All differences are insignificant in comparison to parameterization uncertainties.


\section{Cleo style results of 3-dimensional distribution} \label{sec:3dplots}

In this Appendix we collect 3 dimensional distributions for 3 scalar final 
states following definition used by CLEO \cite{Asner:1999kj}. For each figure, eight 
Dalitz plots in $s_1$, $s_2$ invariant masses are given for $Q^2$ restricted, 
respectively to  0.36-0.81, 0.81-1.0, 1.0-1.21, 1.21-1.44, 1.44-1.69, 1.69-1.96, 1.96-2.25, 2.25-3.24 $\mathrm{GeV^2}$ ranges.

Such representation does not constrain full differential nature of 
hadronic currents, but it was used in publication of experimental data  
of the most differential nature until now. 
It is an important reference point.

\begin{figure}[h!]
\centering
\includegraphics[scale=.650]{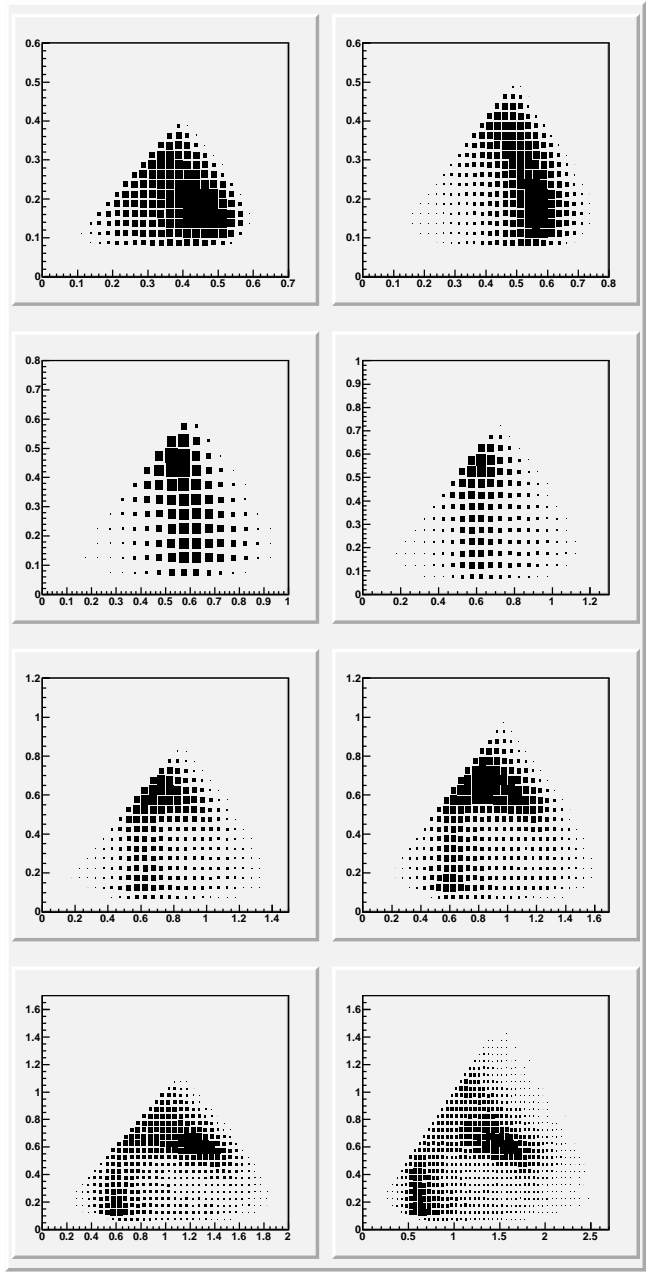}
\caption{8 Dalitz plots for slices in $Q^2$: 
0.36-0.81, 0.81-1.0, 1.0-1.21, 1.21-1.44, 1.44-1.69, 1.69-1.96, 1.96-2.25, 2.25-3.24 $\mathrm{GeV^2}$. 
Each Dalitz plot is distribution for {\bf TAUOLA CLEO} 
in $s_1$, $s_2$ variables ($\mathrm{GeV^2}$ units). 
$s_1$ is taken to be the highest of the two possible values of $M_{\pi^-\pi^+}^2$ in each event. 
\label{fig:CLEOdefault}}
\end{figure}

\begin{figure}[h!]
\centering
\includegraphics[scale=.650]{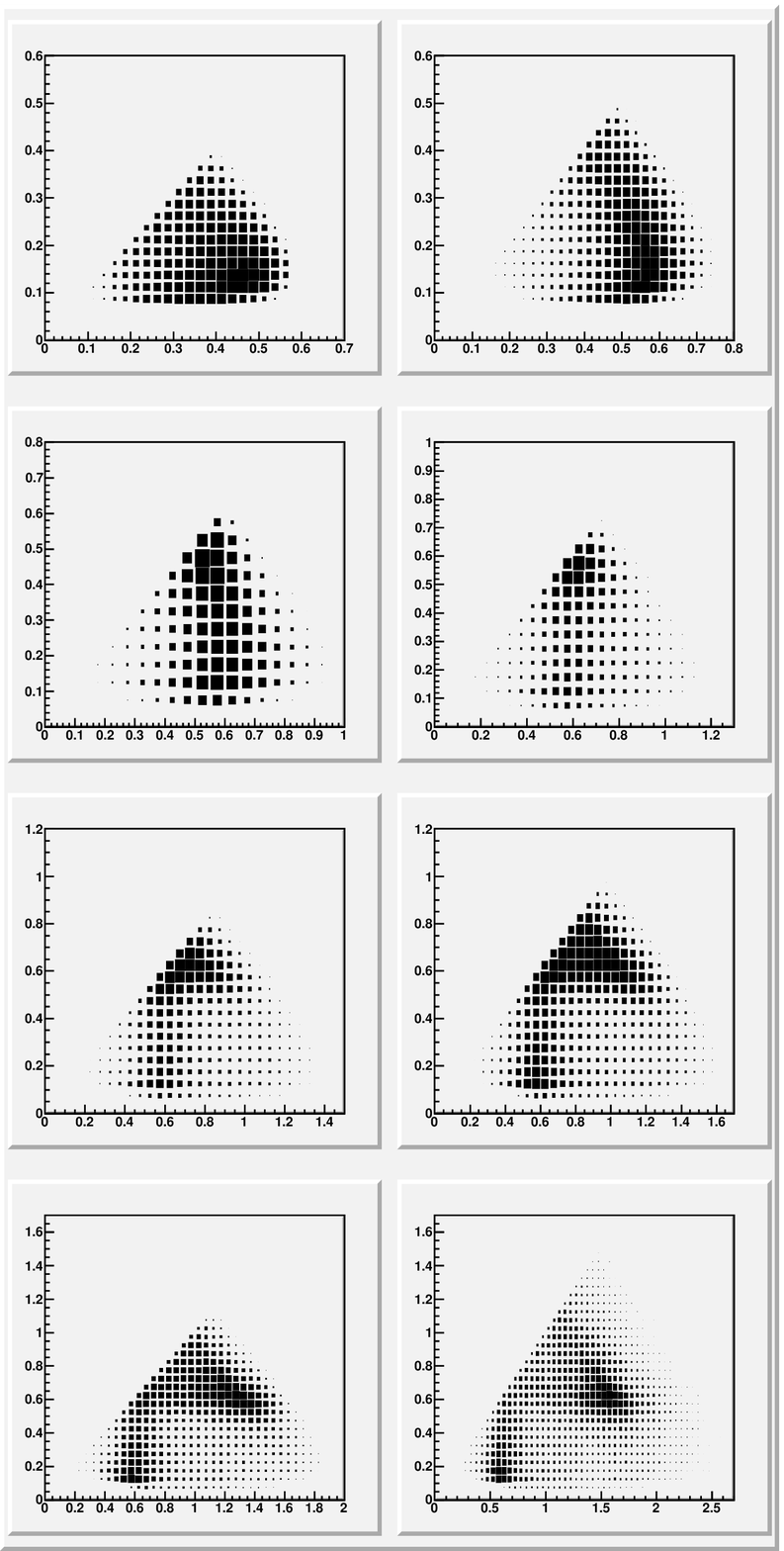}
\caption{8 Dalitz plots for slices in $Q^2$: 
0.36-0.81, 0.81-1.0, 1.0-1.21, 1.21-1.44, 1.44-1.69, 1.69-1.96, 1.96-2.25, 2.25-3.24 $\mathrm{GeV^2}$. 
Each Dalitz plot is distribution for {\bf Pythia CLEO} 
in $s_1$, $s_2$ variables ($\mathrm{GeV^2}$ units). 
$s_1$ is taken to be the highest of the two possible values of $M_{\pi^-\pi^+}^2$ in each event. 
\label{fig:Pythiadefault}}
\end{figure}

\begin{figure}[h!]
\centering
\includegraphics[scale=.650]{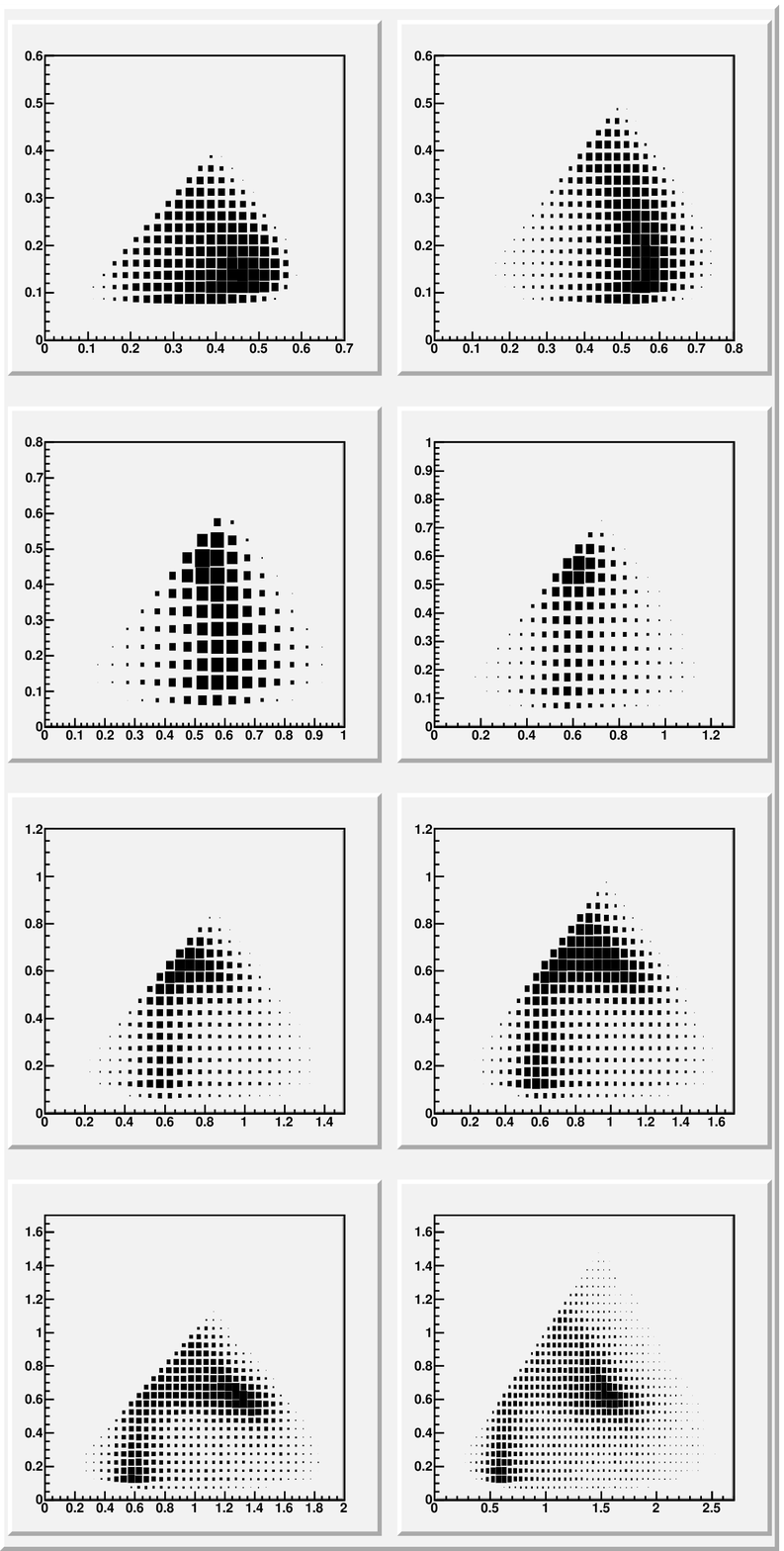}
\caption{8 Dalitz plots for slices in $Q^2$: 
0.36-0.81, 0.81-1.0, 1.0-1.21, 1.21-1.44, 1.44-1.69, 1.69-1.96, 1.96-2.25, 2.25-3.24 $\mathrm{GeV^2}$. 
Each Dalitz plot is distribution for {\bf TAUOLA CLEO isospin intricate} 
in $s_1$, $s_2$ variables ($\mathrm{GeV^2}$ units). 
$s_1$ is taken to be the highest of the two possible values of $M_{\pi^-\pi^+}^2$ in each event. 
\label{fig:CLEOnoIsospin}}
\end{figure}

\begin{figure}[h!]
\centering
\includegraphics[scale=.650]{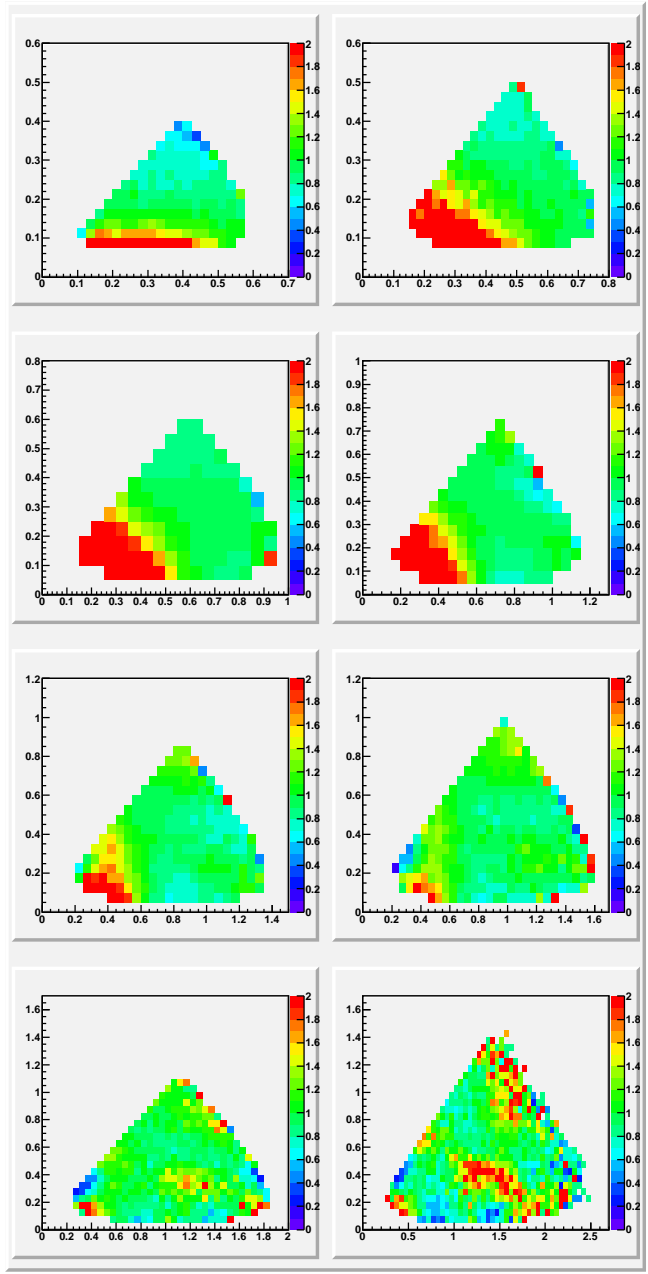}
\caption{
Ratios of {\bf Pythia CLEO} to 
{\bf TAUOLA CLEO} Dalitz plots (Figs.~\ref{fig:Pythiadefault},~\ref{fig:CLEOdefault}).
\label{fig:ratio1}}
\end{figure}

\begin{figure}[h!]
\centering
\includegraphics[scale=.650]{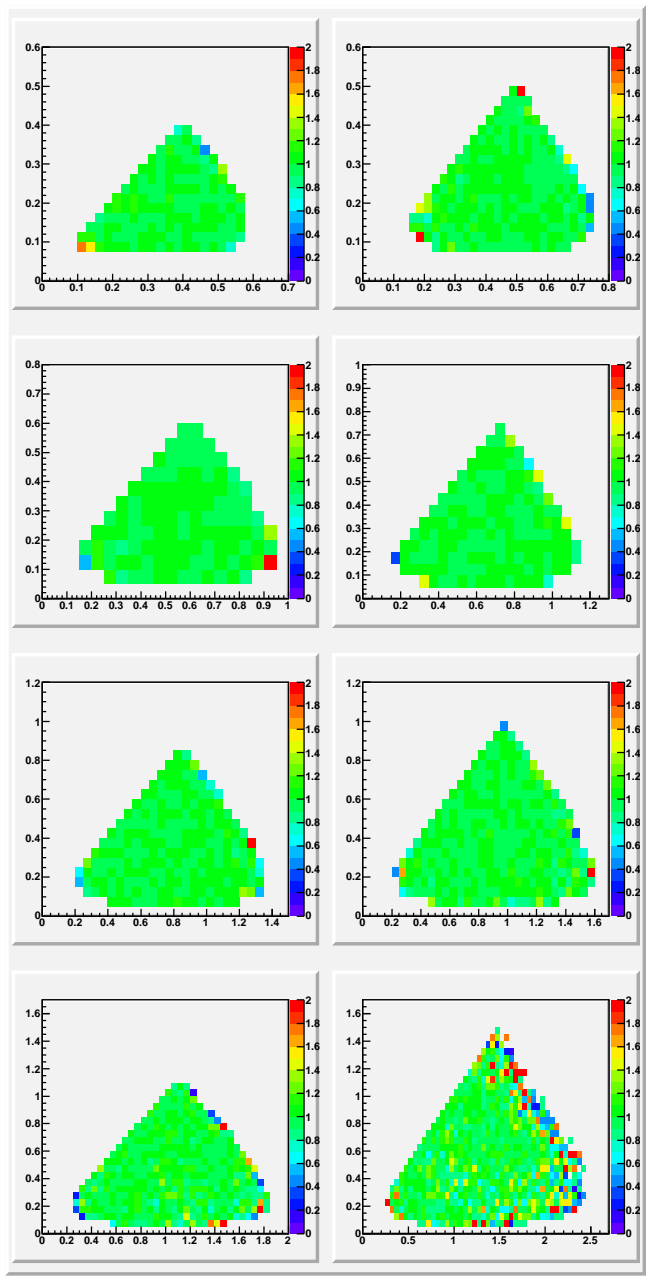}
\caption{
Ratios of {\bf Pythia CLEO} to 
{\bf TAUOLA CLEO isospin intricate} Dalitz plots (Figs.~\ref{fig:Pythiadefault},~\ref{fig:CLEOnoIsospin}).
\label{fig:ratio1a}}
\end{figure}

\begin{figure}[h!]
\centering
\includegraphics[scale=.650]{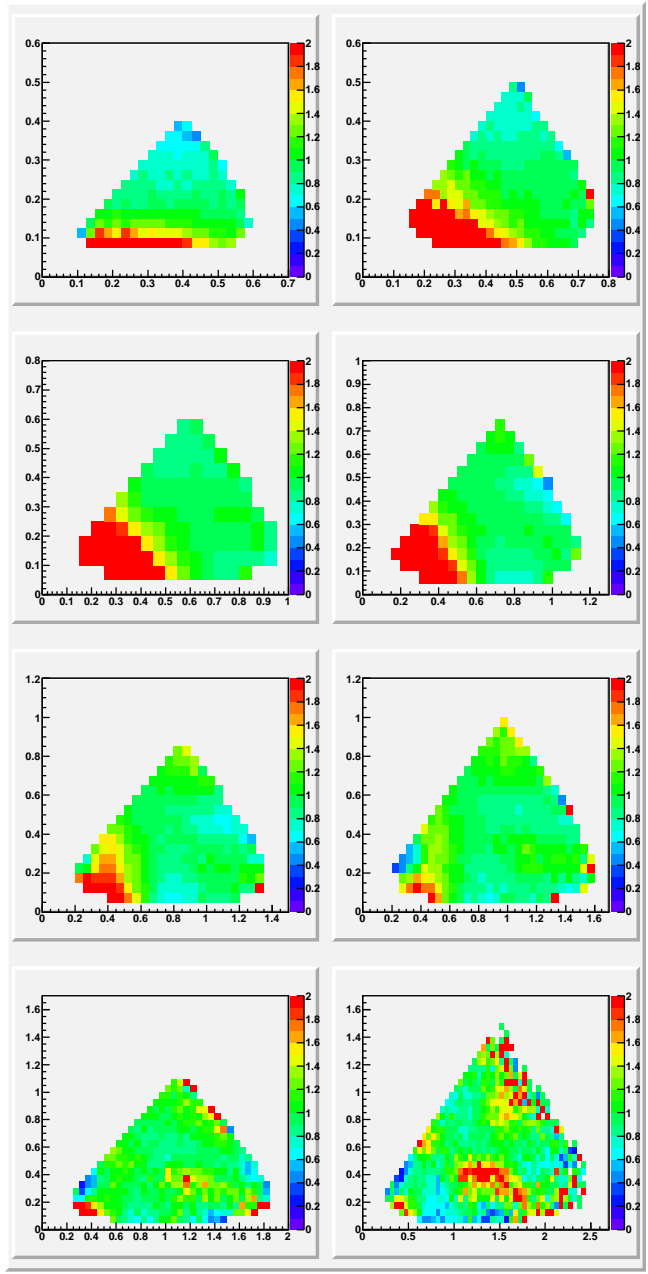}
\caption{
Ratios of {\bf TAUOLA CLEO isospin intricate} to 
{\bf TAUOLA CLEO} Dalitz plots (Figs.~\ref{fig:CLEOnoIsospin},~\ref{fig:CLEOdefault}).
\label{fig:ratio1b}}
\end{figure}

\begin{figure}[h!]
\centering
\includegraphics[scale=.650]{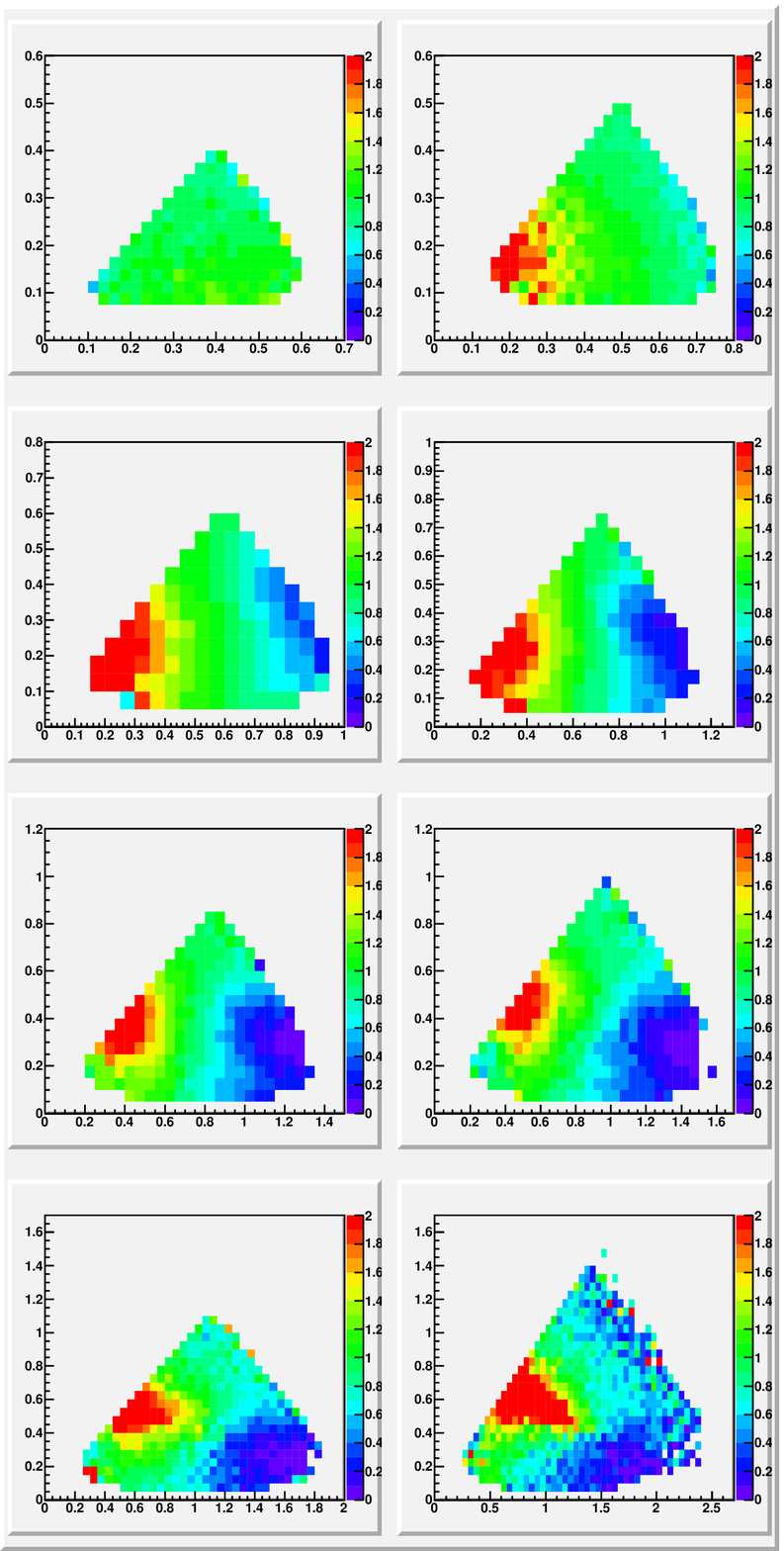}
\caption{Ratios of {\bf TAUOLA BaBar} to {\bf TAUOLA CLEO} Dalitz plots 
in $s_1$, $s_2$ variables ($\mathrm{GeV^2}$ units). Consecutive plots correspond to slices in $Q^2$: 
0.36-0.81, 0.81-1.0, 1.0-1.21, 1.21-1.44, 1.44-1.69, 1.69-1.96, 1.96-2.25, 2.25-3.24 $\mathrm{GeV^2}$.
\label{fig:ratio2}}
\end{figure}

\begin{figure}[h!]
\centering
\includegraphics[scale=.650]{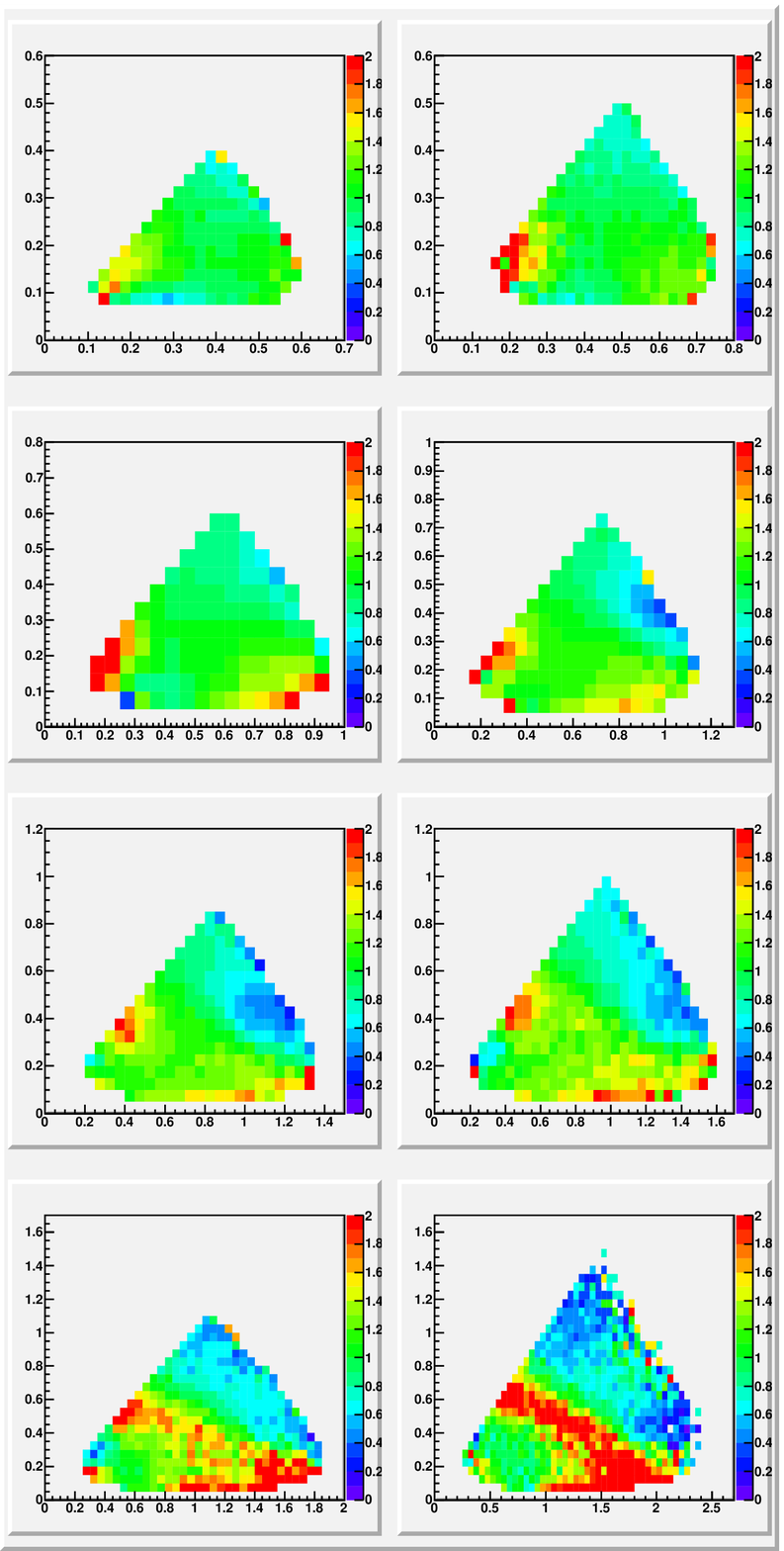}
\caption{Ratios of {\bf TAUOLA RChL} to {\bf TAUOLA CLEO} Dalitz plots 
in $s_1$, $s_2$ variables ($\mathrm{GeV^2}$ units). Consecutive plots correspond to slices in $Q^2$: 
0.36-0.81, 0.81-1.0, 1.0-1.21, 1.21-1.44, 1.44-1.69, 1.69-1.96, 1.96-2.25, 2.25-3.24 $\mathrm{GeV^2}$.
\label{fig:ratio3}}
\end{figure}

\end{document}